\documentstyle[astrobib,psfig]{mn-ab}  

\begin{document}


\newcommand{\PSbox}[3]{\mbox{\rule{0in}{#3}\includegraphics{#1}\hspace{#2}}}

\def\gsim {\lower .1ex\hbox{\rlap{\raise .6ex\hbox{\hskip .3ex
        {\ifmmode{\scriptscriptstyle >}\else
                {$\scriptscriptstyle >$}\fi}}}
        \kern -.4ex{\ifmmode{\scriptscriptstyle \sim}\else
                {$\scriptscriptstyle\sim$}\fi}}}
\def\lsim {\lower .1ex\hbox{\rlap{\raise .6ex\hbox{\hskip .3ex
        {\ifmmode{\scriptscriptstyle <}\else
                {$\scriptscriptstyle <$}\fi}}}
        \kern -.4ex{\ifmmode{\scriptscriptstyle \sim}\else
                {$\scriptscriptstyle\sim$}\fi}}}
\newcommand{\beq}{\begin{equation}}
\newcommand{\eeq}{\end{equation}}
\newcommand{\beqa}{\begin{eqnarray}}
\newcommand{\eeqa}{\end{eqnarray}}
\def\hMpc{h^{-1}{\rm Mpc}}
\def\Msun{M_{\odot}{\ }}
\def\hMsun{h^{-1}M_{\odot}{\ }}
\def\sss{\scriptscriptstyle}
\def\cnfw{c_{\sss \rm NFW}}
\def\rhonfw{\rho_{\sss \rm NFW}}
\def\Mvir{M_{\rm vir}}
\def\mvir{M_{\rm vir}}
\def\Mvirmin{\Mvir^{\rm min}}
\def\Rvir{R_{\rm vir}}
\def\rvir{R_{\rm vir}}
\def\Rvirmin{\rvir^{\rm min}}
\def\Dvir{\Delta_{\rm vir}}
\def\dvir{\Delta_{\rm vir}}
\def\Cvir{c_{\rm vir}}
\def\cvir{c_{\rm vir}}
\def\Rs{r_{\rm s}}
\def\rs{r_{\rm s}}
\def\R2{r_{-2}}
\def\r2{r_{-2}}
\def\rhos{\rho_{\rm s}}
\def\rhost{\tilde\rho_{\rm s}}
\def\Rt{R_{\rm t}}
\def\rt{R_{\rm t}}
\def\Mt{M_{\rm t}}
\def\mt{M_{\rm t}}
\def\Ms{M_{*}}
\def\ms{M_{*}}
\def\rhou{\rho_{\rm u}}
\def\Vvir{V_{\rm vir}}
\def\vvir{V_{\rm vir}}
\def\Vc{V_{\rm c}}
\def\vc{V_{\rm c}}
\def\Vmax{V_{\rm max}}
\def\vmax{V_{\rm max}}
\def\Rmax{r_{\rm max}}
\def\rmax{r_{\rm max}}
\def\Rb{r_{\rm b}}
\def\rb{r_{\rm b}}
\def\ac{a_{\rm c}}
\def\Rd{r_{\rm d}}
\def\rd{r_{\rm d}}
\def\la{\langle} \def\ra{\rangle}
\def\LCDM{$\Lambda$CDM}
\def\lcdm{$\Lambda$CDM}
\def\Omegan{$\Omega_0$}
\def\kms{\ {\rm km\,s^{-1}}}
\def\hmpc{\ h^{-1}{\rm Mpc}}
\def\hkpc{\ h^{-1}{\rm kpc}}
\def\ihmpc{\ h~{\rm Mpc^{-1}}}
\def\Rso{R_{\rm s1}}
\def\Rst{R_{\rm s2}}
\def\Nsp{N_{\rm sp}}
\def\Nspo{N_{\rm sp1}}
\def\Nspt{N_{\rm sp2}}
\def\Mmin{M_{\rm halo}^{\rm min}}
\def\Msp{M_{\rm sp}}
\def\rsp{r_{\rm sp}}
\def\rspo{r_{\rm sp1}}
\def\rspt{r_{\rm sp2}}
\def\Npmin{N_{\rm p}^{\rm min}}
\def\mp{m_{\rm p}}
\def\fres{f_{\rm res}}
\def\xu{\vec{x_u}}
\def\xa{\vec{x_a}}
\def\xb{\vec{x_b}}
\def\ni{\noindent}
\def\omm{\Omega_{\rm m}}
\def\oml{\Omega_{\Lambda}}
\def\zc{z_{\rm c}}

\title[Profiles of dark haloes]
{Profiles of dark haloes: evolution, scatter, and environment}
\author[Bullock et al.]
 {J. S. Bullock$^{1,2}$, T. S. Kolatt$^{1,3}$, Y. Sigad$^3$, 
  R.S. Somerville$^{3,4}$, 
  A. V. Kravtsov$^{2,5}$\thanks{Hubble Fellow.}, \cr
  A. A. Klypin$^5$,
  J. R. Primack$^{1}$, and A. Dekel$^{3}$\\
	$^1$Physics Department, University of California, 
            Santa Cruz, CA 95064 USA\\
	$^2$Department of Astronomy, Ohio State University,
         Columbus, OH 43210 \\
	$^3$Racah Institute of Physics, The Hebrew University, Jerusalem\\
	$^4$Institute of Astronomy, Madingley Rd., Cambridge CB3 OHA, UK \\
	$^5$Astronomy Department, New Mexico Sate University, Box 30001,
            Dept. 4500, Las Cruces, NM 88003-0001 USA} 
\maketitle

\begin{abstract}
We study dark-matter halo density profiles in a high-resolution N-body
simulation  of  a   $\Lambda$CDM  cosmology.   Our statistical  sample
contains  $\sim$$\,5000$  haloes in  the range $10^{11}-10^{14}\hMsun$
and the  resolution allows  a  study of  subhaloes inside host haloes.
The profiles are parameterized by an  NFW form with two parameters, an
inner radius $\rs$ and a virial radius $\rvir$, and we define the halo
concentration $\cvir\equiv\rvir/\rs$.  We  find that, for a given halo
mass, the  redshift dependence of the  median concentration  is $\cvir
\propto (1+z)^{-1}$. This corresponds to $\rs(z)\sim$ constant, and is
contrary to earlier suspicions  that $\cvir$ does  not vary much  with
redshift.  The  implications  are   that high-redshift  galaxies   are
predicted to   be  more extended   and   dimmer than  expected before.
Second, we find  that the scatter in  halo  profiles is large,  with a
$1\sigma$ $\Delta (\log \cvir) = 0.18$  at a given mass, corresponding
to a scatter in maximum  rotation velocities of $\Delta \Vmax/\Vmax  =
0.12$.   We  discuss    implications for  modelling  the  Tully-Fisher
relation, which  has a smaller    reported intrinsic scatter.   Third,
subhaloes  and   haloes   in dense  environments   tend   to   be more
concentrated than isolated  haloes, and show  a larger scatter.  These
results suggest that $\cvir$ is an essential  parameter for the theory
of galaxy modelling,  and   we briefly discuss implications   for  the
universality  of the  Tully-Fisher   relation,  the formation  of  low
surface  brightness galaxies, and the   origin of the Hubble sequence.
We present an improved analytic  treatment of halo formation that fits
the  measured relations between    halo parameters and  their redshift
dependence,   and can   thus  serve  semi-analytic  studies  of galaxy
formation.

\end{abstract} 


\section{INTRODUCTION}
\label{sec:intro}

In the ``standard"    picture of galaxy    formation, dark-matter (DM)
haloes provide  the framework for the  formation  of luminous galaxies
(e.g.,~\citeNP{WR78,BFPR84,WF91}).  The DM  haloes are assumed to form
hierarchically  bottom-up via  gravitational amplification  of initial
density  fluctuations.  The  haloes    carry  with them  gas,    which
eventually  cools and contracts to  form luminous disc galaxies at the
halo   centres.   The halo profile   has  a direct   dynamical role in
determining  the  observable rotation  curve  of  the  disc.  It  also
affects gas cooling and infall and therefore the structural properties
of the  resultant   disc,  such   as   size, luminosity   and  surface
brightness.  In  order  to model properly   the dissipative stages  of
galaxy formation    and obtain meaningful   predictions for observable
quantities   (such as the Tully-Fisher relation),   it is important to
perform detailed dynamical studies of the evolution of halo structure,
and to  obtain statistical characteristics based on  a fair  sample of
the simulated halo population.

Most naturally, the  density profiles of  haloes are expected to be  a
two-parameter family.  This is because, assuming that the formation of
haloes  can be  approximated   by spherical collapse, each  proto-halo
perturbation can be characterized   by two quantities, e.g., mass  and
radius (or density contrast)  at some fiducial  cosmological time.  In
the approximation of  spherical collapse, these parameters specify the
full evolution of each halo, including the epoch at which it collapses
and its virial radius.  A successful two-parameter functional form for
the   halo  profiles has    been    proposed by  Navarro,  Frenk,   \&
White~\citeyear[hereafter NFW95, NFW96, and NFW97]{nfw95,nfw96,nfw97}:
\begin{equation}
\rhonfw(r) = \frac{\rho_s}{(r/\rs)\left(1+r/\rs\right)^2},
\label{eqt:nfw}
\end{equation}
where $\rs$ is  a    characteristic ``inner" radius, and     $\rhos$ a
corresponding  inner density.  As  we show in \S~\ref{sec:NFW}, one of
the inner parameters can be replaced  by a ``virial" parameter, either
the virial radius ($\rvir$), mass ($\Mvir$), or velocity ($\vvir$).  A
very useful alternative  is the concentration parameter $\cvir$, which
relates the inner and virial parameters.
NFW found that this functional form provides a good fit to haloes over
a  large  range  of masses,  and   for several  different cosmological
scenarios.  It has been tested for  the Einstein-deSitter model with a
standard CDM  power spectrum  of  initial fluctuations (SCDM),  a flat
cosmological model with $\omm=0.3$, $\oml=0.7$ and a corresponding CDM
power spectrum ($\Lambda$CDM), and several models with power-law power
spectra  (confirmed by   \citeNP{craig}, and \citeNP{kkk},   hereafter
KKK97).

NFW then  noticed  that, for a  given  cosmology, their  haloes show a
strong correlation   between  the model's  two   parameters, e.g.,  an
increase in $\rho_s$ for decreasing $\Mvir$.  A natural reason for the
fact that low-mass  haloes tend to show  higher densities is that they
typically  collapsed earlier, when  the universe was denser.  To model
this trend, NFW  proposed a toy model  (outlined in Appendix  A) which
assumes  that $\rhos$ is  a   constant multiple  $k$ of the  universal
density $\rhou(\zc)$ at a   collapse   redshift $\zc$, and  that   the
collapsing mass at $\zc$ is a constant fraction  $f$ of the total halo
mass that has   just collapsed and virialized  at  $z=0$.  The general
trend  of the relation between the  two profile parameters at $z=0$ is
reproduced  well for a proper  choice of values  for the constants $k$
and  $f$, with    different  values for  the    different cosmological
models. 

Since the halo profiles are expected  to be a two-parameter family, it
is important to study the scatter about this mean relation between the
two halo parameters.  This scatter  could provide the second parameter
which  is  necessary in  order to explain   the observed variations in
galaxy   properties, such  as bulge-to-disc   ratio, size and  surface
brightness.  It may be argued that the scatter  in halo spin parameter
may also contribute to these variations, but it cannot account for all
of  them  because, e.g., it   does  not  properly correlate  with  the
environment.   The  scatter in halo profiles  should  also have direct
implications for understanding the surprisingly tight scatter observed
in  the Tully-Fisher (TF)  relation  for disc  galaxies.  NFW found  a
small scatter among their  simulated haloes, which could have provided
a   convenient explanation  for  the    small  TF scatter,  but  other
theoretical studies predict a larger scatter~\cite{eisenstein:96}, and
the  generality of the NFW  result is limited   by the small number of
haloes   simulated per cosmology  ($\sim 20$)   and by their selective
choice  of haloes near virial equilibrium.   We  therefore here 
study in detail  the scatter in a large,  ``fair" sample  of simulated
haloes.  A similar investigation has been performed by Jing (2000).

The accumulating  data of galaxies at  high redshifts  provide a great
incentive for   studying the properties of the   halo population  as a
function of redshift.  NFW97 tried to  extend their toy model in order
to predict this redshift  dependence by assuming that  $k$ and $f$ are
both constant  in  time.  In order  to  actually study in  detail  the
redshift dependence of halo   profiles, we use our  large  statistical
sample of simulated haloes. Our  results below, which differ from  the
NFW97 toy-model predictions at $z\geq1$, motivate modifications in the
toy model in order to properly account for the simulated behavior.

We present here an analysis  of a statistical  sample of halo profiles
drawn from cosmological  N-body simulations of $\Lambda$CDM with $\omm
=0.3$  and  $\oml=0.7$.  The  unprecedented  features of this analysis
are:
\begin{itemize}
\item
The sample is large, containing about $5000$ haloes  in the mass range
$10^{11}$--$10^{14} \hMsun$ at $z=0$.
\item
The sampling is ``fair",  in the sense that  haloes  are found in  any
environment, field and  clustered, and  irrespective of the  dynamical
stage of the halo after virialization.
\item
The resolution   is high, allowing  a distinction  between ``distinct"
haloes and ``subhaloes", and a study of environmental trends.
\item
The time evolution and  scatter  about  the one-parameter family   are
studied in detail.
\end{itemize}

In \S~\ref{sec:NFW} we discuss  further the parametric functional form
used for   the halo  profiles.  In   \S~\ref{sec:toy} we  present  the
revised  toy model for predicting the   mean relation between the halo
profile parameters and its  redshift dependence.  In \S~\ref{sec:sims}
we describe  our N-body  simulations 
and
our method of  halo finding and                         
classification, and discuss tests for  the effects of mass  resolution
on  fit parameters.  In \S~\ref{sec:today} we  present our results for
haloes at  $z=0$; we compare the mean  result to our model prediction,
and quantify the  intrinsic  scatter.  In \S~\ref{sec:TF}  we  discuss
implications for observable rotation curves and the TF relation.  In
\S~\ref{sec:redshift}  we investigate the  redshift dependence of halo
properties, and the toy-model fits.  Finally, in \S~\ref{sec:conc}, we
summarize our results and discuss further implications.

\section{PROFILE CHARACTERISTICS} 
\label{sec:NFW}

We choose to fit the  density profiles of all  haloes at all redshifts
with the NFW two-parameter functional  form (Eq. \ref{eqt:nfw}).  This
is a  convenient way  to  parameterize the profiles,  without implying
that   it  necessarily provides  the  best  possible  fit.  A  similar
analysis could be carried out using alternative  functional forms.  In
this section, we  discuss the various  parameters  that are associated
with the halo density profile, the relations between them, and how the
values of these parameters influence observable quantities.

The inner radius, $\rs$,  is where the  effective logarithmic slope of
the profile is $-2$, a characteristic radius which we term $\r2$.  For
much smaller radii, $\rhonfw   \propto  r^{-1}$, and for  much  larger
radii, $\rhonfw \propto  r^{-3}$.  The inner  density parameter of the
NFW  profile is  related to  the  NFW density at   $\rs$  by $\rhos  =
4\rhonfw (\rs)$,  and equals  the local  density at about  half $\rs$:
$\rhos = \rhonfw(r=0.466\,\rs)$.
 
The outer, virial radius $\rvir$, of a halo of virial mass $\Mvir$, is
defined as  the radius within which the  mean density is $\Dvir$ times
the mean  universal   density $\rho_u$   at that  redshift:~\footnote{
$\Mvir  \simeq  10^{11} \hMsun  (\Omega_0  \Dvir(z)/200)[\Rvir(1+z)/75
\hkpc]^3.$}
\beq
\Mvir  \equiv  \frac{4 \pi}{3} \Dvir \rho_u \rvir^3.
\label{eqt:mvir}
\eeq
The associated virial  velocity is defined by\footnote{  
$\Vvir \simeq 75{\rm   km/s}    (\Rvir/75  \hkpc)   
(\Omega_0    \Dvir(z)/200)^{1/3} (1+z)^{3/2}$.}   
$\Vvir^2   \equiv     G\Mvir/\rvir$.   The one-to-one
relations between the three virial  parameters are fully determined by
the background cosmology (independent of the inner halo structure), so
only one  of  them at  a time  can  serve in the  pair  of independent
parameters characterizing the profile.  The virial overdensity $\Dvir$
is provided by the  dissipationless  spherical top-hat collapse  model
\cite{peebles:84,eke:98}; it is a function of  the cosmological model, and it
may vary with time.  For the Einstein-deSitter cosmology, the familiar
value is  $\Dvir \simeq 178$  at all  times.   For  the family  of  flat
cosmologies ($\omm+\oml=1$), the value of  $\Dvir$ can be approximated
by (\citeNP{bryan:97})   
$\Dvir  \simeq   (18\pi^2     +  82x  - 39x^2)/\Omega(z)$, 
where $x\equiv \Omega(z) - 1$, and $\Omega(z)$ is the
ratio of mean matter density to critical density at redshift $z$.  For
example, in the \lcdm\ cosmological model that serves as the basis for
our  analysis in  this paper  ($\omm = 0.3$),   the value at $z=0$  is
$\Dvir\simeq 337$.

An  associated useful characteristic  is the  concentration parameter,
$\cvir$, defined as the ratio between the virial and inner radii,
\beq
\cvir \equiv \rvir / \rs.
\label{eqt:cvir}
\eeq
Note  that our   definition   of $\cvir$ differs  slightly  from  that
originally used by NFW, $\cnfw \equiv R_{200}  / \rs$, where $R_{200}$
is  the radius corresponding  to a density of   200 times the critical
density,  independent  of the  actual cosmological model.

\begin{figure}
\PSbox{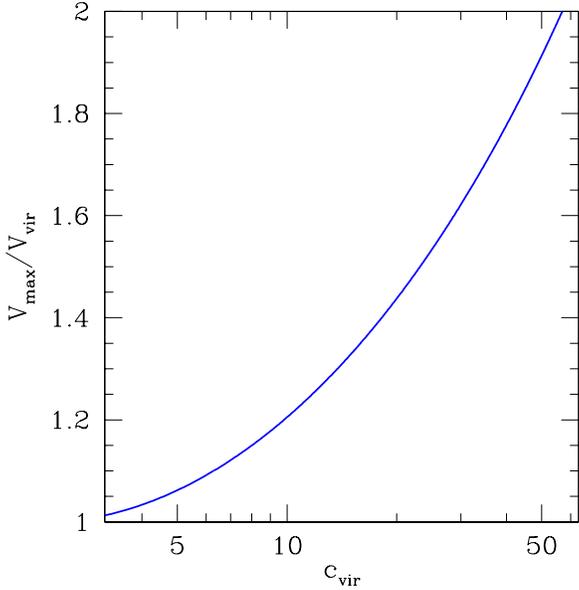  hoffset=-10 voffset=-65 hscale=40 vscale =
40}{74mm}{74mm}

\caption{Maximum velocity versus concentration.
The maximum rotation velocity for an NFW halo in units of the rotation
velocity at its virial radius as a function of halo concentration.  }
\label{fig:fit.check1}
\end{figure}

A third relation between the parameters of the NFW profile is
\beq
\Mvir = 4\pi \rhos \rs^3 A(\cvir), \quad
A(\cvir) \equiv \ln(1+\cvir) - {\cvir \over 1+\cvir}.
\label{eqt:mcvir}
\eeq
The   three  relations    (Eqs.  \ref{eqt:mvir},  \ref{eqt:cvir}   and
\ref{eqt:mcvir})  allow the  usage of any  pair  out of the parameters
defined so far (excluding  degenerate pairs consisting only of  virial
quantities)    as  the   two  independent  parameters that  fully
characterize the profile.

Since the more observable quantities have  to do with rotation curves,
it is  worth translating the density profile  into a circular velocity
curve for the halo,
\beq
\Vc^2(r) 
\equiv {{G M(r)} \over r }
= \Vvir^2 {\cvir \over A(\cvir)} {A(x) \over x } ,
\label{eqt:vmax.of.c}
\eeq
where $x\equiv r/\Rs$.  The maximum velocity occurs at a radius $\Rmax
\simeq 2.16\,\rs$ and is given by
\beq
{\vmax^2 \over \vvir^2} \simeq 0.216\, {\cvir \over A(\cvir)}.
\label{eqt:vmax}
\eeq
For typical $\cvir$ values in the range $5-30$,  $\Vmax$ varies in the
range $(1-1.6)\Vvir$.  Figure  1  shows the  ratio $\Vmax/\Vvir$ as  a
function of $\cvir$.  Note that for haloes of  the same mass, a larger
$\cvir$ goes  with  a larger $\Vmax$.    Because of  the  relationship
between $\Rmax$ and $\rs$, haloes  with $\cvir \lsim 10$ have velocity
curves which continue to rise gradually out to an appreciable fraction
of their virial  radii, while those with  $\cvir  \gsim 10$ rise  more
steeply,   and  possibly  represent   galaxies   in  which $\Rmax$  is
identifiable   observationally   (though   the  effects  of   baryonic
contraction  should  be taken into   account  before the observed  and
simulated values of $\Rmax$ can be compared).

In  order    to gain a   qualitative    understanding of  how  profile
characteristics may affect galactic disc formation, we may assume that
an exponential disc  forms by the  adiabatic contraction of gas inside
the dark-matter halo  \cite{blumenthal,FPBF93,dalcanton,mo:97}.    The
final disc size,  $\rd$,  can  be derived   from the halo   parameters
$\cvir$ and $\Rvir$, under the following  further assumptions that (a)
the   disc forms from   cold  gas   of mass $\sim$0.03$\Mvir$    (see,
e.g. \citeNP{sp:99}) which follows the original density profile of the
halo out to $\Rvir$, and (b) the  specific angular momentum of the gas
is equal to that of the halo, which has  an original spin parameter of
$\lambda = 0.035$ ---  the most probable value  of the  well-known log
normal distribution.  Under these assumptions,  we find the  following
fitting formula is good to within $1 \%$ for $1 < \cvir < 50$
\cite{bullock:c}:
\beq
\rd \simeq 5.7 \hkpc \left(\frac{\Rvir}{100 \hkpc }\right) 
[1 + (\cvir/3.73)^{0.71}]^{-1}.
\label{eqt:rd}
\eeq
(A similar fitting formula, which allows more varied assumptions about
the halo   and  disc make-up, is   provided  by  \citeNP{mo:97}.)  The
general result is that $\rd$, and thus the disc surface brightness, is
a decreasing function of $\cvir$.

We  stress again that  the specific choice  of the NFW functional form
does not limit the generality of our  analysis in a  severe way.  This
is largely due to the association of the  specific $\rs$ with the more
general $\r2$. When the NFW function is fitted to a generic halo whose
profile even vaguely resembles a similar  shape, the fitting procedure
is  likely to return an  $\rs$  value that  is  close to the effective
$\r2$  of that   halo.     The concentration  parameter   can  then be
interpreted  as a   general    structure parameter   not   necessarily
restricted to the specific NFW function.  In particular, any spread in
$\cvir$ can be  attributed to a  real scatter  in a  ``physical" inner
radius such   as $\r2$  rather than  to  inaccuracies  in the  assumed
universal profile.

The   interpretation of  $\rs$ as  $\r2$  allows  one to  map the  NFW
parameters  to appropriate parameters of other  functional forms.  For
many  purposes, such as determining  $\Vmax$ or  modelling gas cooling
and galaxy formation, the NFW form is sufficiently accurate.  However,
a comparison with alternative profiles  with different core  behaviors
may be important when much smaller radii are  concerned ($r \lsim 0.02
\rvir$),  where there   are   indications   of  deviations from     an
extrapolation of the  NFW shape (\citeNP{kravtsov:98}, hereafter KKBP98;
\citeNP{joel.rutgers,moore}; \citeNP{kkbp:00}, hereafter KKBP00).

A specific example of an alternative profile functional form is the
Burkert profile \citeyear{burkert}:
\beq
\rho_{\sss \rm B}(r) = \frac{\rho_{\rm b}}
     { [1 + (r/\Rb)^2 ] [1+r/\Rb]}.
\eeq
This profile resembles the NFW profile for  $r \gsim 0.02 \rvir$.  The
Burkert profile has a log slope of $-2$ at $\r2 \simeq 1.52\, \Rb$, so
one can  relate  the scale radii  of  NFW and  Burkert by  $\Rb \simeq
\rs/1.52$, and then relate  the concentration parameters by $\rvir/\Rb
\simeq 1.52\, \cvir$.

Note also that the relationship between $\cvir$ and $\Rmax$ is robust,
regardless of profile shape.  For example, with the Burkert profile we
have an implied velocity maximum at $\Rmax \simeq 3.25\, \Rb$.  If one
assumes instead the relation  as gleaned from  an NFW fit, with $\Rmax
\simeq 2.16 \rs \simeq 3.28 \Rb$, there  is good agreement between the
values of $\Rmax$.

\section{A REVISED TOY MODEL}
\label{sec:toy}

In any  investigation  based on computer  simulations  it is useful to
have a simple toy model that helps interpret the numerical results and
allows  an easy application of the  conclusions in subsequent analytic
or semi-analytic investigations.  As mentioned in the Introduction and
outlined in Appendix A, NFW96 and NFW97  proposed such a model, with 2
free parameters,  which successfully recovers the mean $\cvir$-$\mvir$
relation at  $z=0$  for  the  several different   cosmological  models
simulated by them.  Our simulations (see \S5 below) indeed confirm the
success    of   this   model   at     $z=0$.   However,     we    find
(\S~\ref{sec:redshift} below) that  the NFW97 model does not reproduce
properly the redshift dependence of  the halo profiles  as seen in the
simulation; it significantly over-predicts the concentration of haloes
at early times, $z \gsim 1$.  We therefore propose a revised toy model
that is shown below to recover properly the full  behavior of the mean
$\cvir$-$\mvir$ relation and  its redshift dependence.  We present the
model in this preparatory section, so that we can refer  to it when we
describe   and  interpret the results    from  the simulations  in the
following sections.  A fortran code that implements this model for 
several cosmologies is available from the first author upon
request.~\footnote{http://www.astronomy.ohio-state.edu/~james/CVIR/parts.html}

\subsection{The model}

We seek  a model for the typical  halo concentration, denoted  in this
section by $\cvir(\mvir,a)$,  for a given  mass $\mvir$ and epoch 
$a = (1+z)^{-1}$.  Following the general spirit of the NFW97 model, we
assign  to  each halo  an  epoch of    collapse, $\ac$.   Unlike their
formulation,    which  utilizes     the      extended  Press-Schechter
approximation, we define $\ac$ in a simpler  way as the epoch at which
the typical collapsing mass,  $\Ms(\ac)$, equals a fixed  fraction $F$
of the halo mass at epoch $a$,
\beq
\Ms(\ac) \equiv F \Mvir.
\label{eqt:toy1}
\eeq
The typical collapsing  mass  at an  epoch $a$ is  defined by  
$\sigma [\Ms(a)]=1.686/D(a)$,
where $\sigma  (M)$  is the $a=1$ linear  rms  density  fluctuation  on the
comoving scale encompassing  a  mass $M$, $D(a)$ is the linear
growth rate,   
and $1.686$ is   the linear
equivalent   of the density  at  collapse   according to the  familiar
spherical collapse model.  The typical collapsing  mass is therefore a
known  function of the linear power   spectrum of fluctuations and 
$D(a)$ for the cosmology in hand.
For some purposes we  find it convenient to measure  the halo mass  in
units of the typical halo mass at the same epoch,
\beq
\mu \equiv \mvir(a)/ \Ms(a).
\label{eqt:mu}
\eeq

The second relation of the model arises by  associating density of the
universe $\rhou$ at $\ac$ with a characteristic density of the halo at
$a$.  NFW97 used the inner density parameter $\rhos$, which is related
to $\mvir$ and   $\rs$ via the   specific  shape of the  NFW  profile.
Instead,  we define a more general  characteristic density $\rhost$ by
combining inner and virial quantities:
\beq
\mvir \equiv {4 \pi \over 3} \rs^3 \rhost.
\label{eqt:toy2}
\eeq
The NFW profile implies $\rhost =  3 \rhos A(\cvir)$.  The association
of the halo density $\rhost$ with the universal density at collapse is
made via a second free parameter, $K$:
\beq
\tilde{\rho_s} =  K^3 \Dvir(a) \rho_{u}(\ac) 
    =  K^3 \Dvir(a) \rho_{u}(a) \left({a\over\ac}\right)^3.
\label{eqt:toy3}
\eeq
The parameter $K$ represents contraction of the inner halo beyond that
required  for  top-hat dissipationless  halo virialization,  and it is
assumed   to       be  the    same    for     all      haloes.   Using
Eqs.~\ref{eqt:mvir},~\ref{eqt:toy2}, and~\ref{eqt:toy3}, we  obtain  a
simple expression  for $\cvir$ in terms of  $\ac$ as our  second model
relation:
\beq
     \cvir(\mu,a) = K  {a\over\ac} . 
\label{eqt:toy4}
\eeq

The     model  is      fully   determined    by    Eqs.~\ref{eqt:toy1}
and~\ref{eqt:toy4} given the values of the two parameters $F$ and $K$.
We find below  (\S~\ref{sec:today} and  \S~\ref{sec:redshift}) that by
adjusting $F$ and $K$ we  are able to reproduce  the full behavior  of
$\cvir(\mu,a)$ as measured in our  simulations.  The small differences
in the definitions of $F$ and $K$ compared to the analogous parameters
of NFW97, $f$ and $k$,  make a big   difference in the success of  the
model.

Note in Eq.~\ref{eqt:toy1} that, for  any cosmology, $\ac$ is uniquely
determined  by  $\Mvir$,   independent   of $a$.   This  implies   via
Eq.~\ref{eqt:toy4} that, for a fixed halo mass,
\beq
\cvir(a) \propto a .
\label{eqt:cwz}
\eeq
Our   model thus predicts    that  for haloes  of   the same  mass the
concentration is proportional to $(1+z)^{-1}$.  This is different from
the  NFW  prediction  in  which  the concentration  is  a much  weaker
function of redshift.

In order to  gain a basic understanding of  the important  elements of
this model,   we discuss its    predictions in the context   of  three
cosmological models of increasing complexity: (1) a self-similar model
of Einstein-deSitter  cosmology  and a   power-law power  spectrum  of
fluctuations, (2)  standard   CDM,  in  which the  universe   is still
Einstein-deSitter but  the power spectrum  deviates from  a power law,
and  (3) the   relevant   cosmology  of the  current    investigation,
$\Lambda$CDM, with $\omm\neq 1$ {\it and} a non-power-law spectrum.

\subsection{Example 1: the self-similar case}

As an  illustrative   example,   assume a  fully   self-similar  case:
Einstein-deSitter cosmology, $\omm = 1$, for which  the growth rate is
$D(a)\propto a$, and a power-low power spectrum of fluctuations,
$P(k)\propto k^n$, 
for which $\sigma  (M) \propto M^{-\alpha}$  with $\alpha =  (n+3)/6$.
In this case,
\beq
\Ms(a) \propto a^{1/\alpha} .
\eeq
Together with Eqs.~\ref{eqt:toy1} and \ref{eqt:mu}, we have
$\ac/a_0 = (\mu F)^{\alpha}$. 
Then, using Eq.~\ref{eqt:toy4}, we obtain
\beq
 \cvir(\mu,a) = K (F\mu)^{- \alpha}.
\label{eqt:toy6}
\eeq

Note that in the self-similar case the two  parameters $F$ and $K$ can
be replaced by one parameter,  $K F^{-\alpha}$.  Equivalently, we  may
vary only  $K$ and adopt  the natural  value $F=1$,  namely, apply the
model to the collapse of the whole halo.  This is a special feature of
our revised model, not valid in the original NFW model.

The slope of the function $\cvir(\mu)$ at $a$ is completely determined
by the power index $\alpha$ (i.e., $n$): $\cvir \propto \mu^{-\alpha}$.  
This simple mass dependence can be checked  against the results of the
simulations of NFW97 in their Figure 6, which presents $\cvir(\mu)$ at
$z=0$ for Einstein-deSitter cosmology   with four different  power-law
spectra: $n=-1.5,  -1, -0.5$, and $0$  ($\alpha = 1/4,  1/3, 5/11$ and
$1/2$).  In  each case, our model  predicts  the simulated slope quite
well, even  slightly  better than the   NFW model, 
and, unlike the NFW model,
 it predicts an exact power law relation between $\cvir$ and
$\mu$, as would be expected for a scale-free cosmology.
   As  the power-law
$M^{-\alpha}$ becomes steeper ($n$ larger), the difference in collapse
epochs for haloes of a given mass difference  becomes larger, which is
reflected in a steeper $\cvir(\mu)$ relation.

The collapse factor $K$ may be determined  from the simulations at the
present epoch by  matching the value of $\cvir$  at any desired $\mu$.
We find  for a typical simulated  halo ($\mu=1$) at $a=1$  that $\cvir
\simeq 10$, implying that the additional collapse factor for the whole
halo ($F=1$) must be $K \sim 10$.

Clearly,  for  the  fully   self-similar  case one  would   expect any
dimensionless  properties of  $\Ms$ haloes  to  be invariant  in time.
This is  easily verified by setting  $\mu = 1$  in Eq.~\ref{eqt:toy6}:
$\cvir(1,a) = K$ (for $F=1$).  In this case, the concentration is also
fixed for any other fixed value of $\mu$.
Accordingly, the concentrations are  different for haloes of  the {\it
same  mass} that  are  addressed  at   different redshifts.   For  the
Einstein-deSitter cosmology,  Eq.~\ref{eqt:mvir} yields 
$\rvir \propto a$,
and then the general behavior of $\cvir \propto a$ (Eq.~\ref{eqt:cwz})
implies (via Eq.~\ref{eqt:cvir}) that the  value of $\rs$ is the  same
(in physical coordinates) at all  redshifts.  Again, this is different
from the NFW97 model prediction.

\subsection{Example 2: SCDM}

As a second example, consider the  standard CDM cosmology (SCDM: $\omm
= 1$, $h=0.5$,  $\sigma_8=0.7$).  Although the  time evolution here is
still self-similar, the power spectrum is not a power law: $\sigma(M)$
has  a characteristic bend near $\mu  \sim 1$ today [which corresponds
to a  mass  $\Ms(a=1) \simeq 3.5\times  10^{13}   \hMsun$].  The local
slope varies from $\alpha \simeq 0$ for $\mu  \ll 1$ to $\alpha = 2/3$
for $\mu \gg 1$.

The  model solution for  $\cvir(\mu)$ does not  have  a closed form in
this case, but it is easy to obtain a useful approximation.  The slope
of the $\cvir(\mu)$ relation at a specific  $\mu$ is determined by the
effective slope of the power spectrum on a  scale corresponding to the
mass $\mu F \Ms$ (not $\mu \Ms$), since this is the mass scale used to
characterize   the  halo  collapse  time  in   Eqs.~\ref{eqt:toy1} and
\ref{eqt:mu}).  We now   obtain an  approximate relation  similar   to
Eq.~\ref{eqt:toy6},  but  with the effective local  $\alpha$ replacing
the constant $\alpha$:
\beq
 \cvir \simeq K (\mu F)^{- \alpha (\mu F)}.
\label{eqt:toy7}
\eeq
Unlike the power-law example,  the value chosen  for the  constant $F$
does play a role in the slope of the  $\cvir(\mu)$ relation at a given
$a$.  In order  to determine the best values  of $F$ and $K$, we match
the model predictions  of $\cvir(\mu)$ to the  results of the $N-$body
simulations of  the SCDM model at  the  present epoch.  Using $F=0.01$
and $K=4.5$ in Eqs.~\ref{eqt:toy1}  and \ref{eqt:toy4}, we are able to
reproduce quite well the $z=0$ SCDM results of  NFW97 (their Figure 6)
over the range $\mu \simeq 0.01 - 100$.  The  relation about $\mu \sim
1$ is well  approximated using  Eq.~\ref{eqt:toy7}, where  $\alpha(\mu
F=0.01) \simeq 0.14$.

But now, using no extra parameters, we are also  able to reproduce the
time dependence  of  this relation, which  we  test using the  $\cvir$
values as determined from our small box (7.5 $\hMpc$) SCDM simulation.
In our simulation,  we find consistent results with the NFW97
simulation at $z=0$. In addition, for  a
fixed $\Mvir$, we find  that the model-predicted scaling  of $\cvir(a)
\propto a$  indeed describes very well the  time evolution of the halo
population.

In   our  toy model,  the  values  of $F$  and $K$   are assumed to be
constants  as a function of both  $a$ and $\mu$.    Such a behavior is
naturally expected in the fully self-similar case, in which no special
time or scale is present in the problem.  However, the success of this
toy model in the SCDM case, which is  not scale invariant, is somewhat
surprising.  The reason for this success is  linked to the small value
of $F$, which pushes the relevant mass scales of the problem to values
much below that of the bend in the power spectrum (near $\mu \sim 1$),
where  the spectrum is   approaching a power-law behavior.  For  small
values of $\mu$ ($\mu \lsim 0.1$), the slope of $\cvir(\mu)$ is almost
independent  of the actual value   of $F$ as  long  as  the latter  is
smaller than  $0.05$.   The specific  preferred  value  of $F =  0.01$
arises from  the  need  to   match the  model $\cvir(\mu)$  with   the
simulations in the range $\mu \geq 1$.

\subsection{Example 3: $\Lambda$CDM}

Our third  example    concerns  a   currently   popular   $\Lambda$CDM
cosmological model  ($\omm=0.3$, $\oml=0.7$,  $h=0.7$, $\sigma_8=1.0$,
where  $\Ms \simeq 1.5\times10^{13} \hMsun$  at  $z=0$).  In this  case,
self  similarity is  violated  due to  the  non-power-law spectrum  as
before, and  also   by  the time-dependent  fluctuation   growth  rate
associated  with the low  $\omm$.     Using similar reasoning to   our
discussion in the previous example, one  may expect our toy model with
constant $F$ and $K$ to break down.    These worries are
somewhat alleviated as  long as $F$ is  small.   As in  the SCDM case,
this pushes the relevant mass scales to the power-law regime away from
the bend  in the power spectrum.   In  addition, a small value  of $F$
demands that the collapse  epoch is early,  when the mean  density was
near   the critical value,  $\omm(\ac)  \simeq  1$,  and therefore the
fluctuation   growth rate  was  close  to  that  in  the  self-similar
cosmology,
$D(a) \propto a$.  
The deviation   from  the  self-similar   growth rate   introduces  in
Eq.~\ref{eqt:toy7}  a  multiplicative correction  factor, given by the
growth of fluctuations  between the epochs  $\ac$ and $a$ in the given
cosmology   compared to the Einstein-deSitter case.    In  the case of
$\Lambda$CDM and $\ac\ll 1$ this factor at $a=1$ is about $1.25$.

Although the small value of  $F$ alleviates most the expected problems
associated with the model, for open cosmologies,
there will be a  break   down  at  extremely     high masses   
$M  \gg F^{-1}M_*(a=1)$.    
This is because  the fluctuation growth  rate saturates at late times,
$D(a \gg  1)   \rightarrow$  const., and  the    definition of collapse
redshift  (Eq.~\ref{eqt:toy1})    loses meaning.   For our best-fit
value, $F=0.01$, this limits the applicability of our  model to haloes less
massive than $\sim 100   M_*(a=1)$.  Fortunately, haloes more massive than
this are extremely rare (by definition) so the range of applicability
covers almost all relevant cases.~\footnote{Choosing $F=0.001$ and $K=3.0$ 
extends the applicable range of our model to $\lsim 1000  M_*(a=1)$,
and results in only a slight over-prediction of $\cvir$ values
compared to the $M \sim 10^{14} \hMsun$ haloes in our simulation,
and a better fit to the $\Lambda$CDM haloes of NFW97.}

As   we show in    \S~\ref{sec:today} and \S~\ref{sec:redshift} below,
using $F=0.01$ and  $K=4.0$ in the  model equations \ref{eqt:toy1} and
\ref{eqt:toy4}, we are  able to  reproduce the  full behavior  of  the
median  $\cvir(\mu,a)$ in our $\Lambda$CDM  simulations.  At $a=1$, we
have $\cvir(\mu) \propto  \mu^{-\alpha(F \mu)}$,  where 
$\alpha \simeq 0.13$ for $\mu \sim 1$.  For haloes of fixed mass 
($\mu\Ms = {\rm const}$), we have $\cvir(a) \propto a$ as before.  The
same  model  parameters also provide  a  reasonable  fit to the  $z=0$
$\Lambda$CDM results of NFW97 over  the range $0.01  M_* \lsim M \lsim
100 M_*$.

We will show below that by setting the value of $K$ to  2.6 and 6.0 we
are able to  artificially parameterize  the $-1\sigma$ and  $+1\sigma$
scatter  respectively in  the  value of   $\cvir$  for the   simulated
population of haloes.  A similar range  of $K$ values accounts for the
scatter for all masses and at all cosmological epochs.

\section{SIMULATING HALOES}  
\label{sec:sims}

\subsection{The numerical simulations}
\label{sec.sims.haloes:sims}
                                                             
Only recently  have large cosmological  N-body simulations reached the
stage where detailed structural  properties of many dark-matter haloes
can  be resolved simultaneously.  One  of  the most successful methods
for high  force    resolution and mass   resolution is   the  Adaptive
Refinement Tree  (ART)   code  (KKK97).   The method   makes   use  of
successive refinements of   the grid and  time  step  in high  density
environments.  The simulations based on  the ART code provide, for the
first time, a compilation of a statistical  sample of well-resolved DM
haloes, as well as substructure  of haloes within haloes.  In previous
simulations,  haloes  were picked ``by  hand"  using certain selection
criteria  from a   low-resolution  cosmological  simulations,   to  be
re-simulated with high resolution.  This  selection induces a  certain
bias into the sample.

We have  used the  ART code to   simulate  the evolution   of DM  in a
low-density flat $\Lambda$CDM model  for which $\omm=0.3$, $\oml=0.7$,
$h=0.7$, and $\sigma_8=1.0$   at $z=0$.  The  simulation  followed the
trajectories   of $256^3$ cold dark  matter  particles within a cubic,
periodic box of comoving size $60\hmpc$ from redshift  $z = 40$ to the
present.   We  have  used  a $512^3$    uniform grid,  and up   to six
refinement levels in    the regions of  highest  density,   implying a
dynamic range of $32,768$.  The formal resolution of the simulation is
thus  $\fres = 1.8\hkpc$, and  the mass per  particle is  $m_{p} = 1.1
\times 10^{9} \hMsun.$ In the present paper,  we analyze 12 saved time
steps from $z=5$ to $0$.  We have also used two simulations in smaller
boxes to check for resolution and cosmology  dependence.  One of these
is a $30\hmpc$ box simulation of the same $\Lambda$CDM cosmology, with
$256^3$  particles, $m_{p} = 1.4 \times  10^{8} \hMsun$,  and $\fres =
0.9\hkpc$.  The other,  in a $7.5\hmpc$ box,  is of the SCDM cosmology
($\omm=1$,  $h=0.5$, and $\sigma_8=0.7$  at $z=0$), and it consists of
$128^3$ particles, $\fres = 0.5\hkpc$, and $m_{p}  = 5.5 \times 10^{7}
\hMsun$.  Tests of the ART code  for numerical effects on halo density
profiles are discussed in KKBP98 and KKBP00.

\subsection{Finding and fitting haloes and subhaloes} 
\label{sec:sims:finding}
                                                                     
In this investigation, we sample all types of DM haloes independent of
their environment.  In particular, we identify  both the standard kind
of    ``distinct"    haloes, of the     type  identified  using common
friends-of-friends   algorithms   and  considered  in  Press-Schechter
approximations, and also ``subhaloes",    whose centres  are   located
within  the   virial radius  of  a  larger  ``host''  halo.   Our halo
finding/classifying algorithm,  which is  based  on the Bound  Density
Maxima technique (\citeNP{kh}), has    been specifically designed   to
simultaneously identify distinct haloes and subhaloes (Appendix B).

We   fit   every   identified  DM   halo     using   the NFW   profile
(Eq.~\ref{eqt:nfw}).  Before fitting, we check the halo radial density
profile to see if it has a significant upturn, $d\rho(r)/d r > 0$, and
if so, we  declare this point to  be the truncation radius $\rt$.  Our
measured $\rt$   values are comparable  or  somewhat smaller  than the
expected  tidal  radii.  For  haloes  with  no significant  upturn  in
density, we  fit the NFW density  profile using logarithmically spaced
radial  bins from  0.02$\rvir$ to $\rvir$,   while for  haloes  with a
truncation radius,  we fit the  profile  from 0.02$\rt$ to  $\rt$, and
extrapolate the NFW function  in order  to  assign fitted $\rvir$  and
$\Mvir$.

The profile fitting   is performed as  follows.    After identifying a
centre for the halo, we count particles  in each radial bin and assign
corresponding Poisson errors based on the count  in each bin.  We then
fit  an NFW profile (by  $\chi^2$ minimization) to  the counts in bins
using  the  bin errors in  the  covariance matrix, and obtain best-fit
values for the two free  parameters $\rvir$ and $\rs$ (or equivalently
$\Mvir$  and $\cvir$, etc.)  along  with the  corresponding  errors in
these parameters.  We then remove unbound particles from each halo, as
described in Appendix B, and iterate the  process of determining $\rt$
and fitting a profile until the halo contains only bound particles.

We    present    results  for  haloes   with     masses in   the range
$1.5\times10^{11} - 10^{14} \hMsun$; the  smallest haloes thus contain
$\gsim 150$ particles.   A profile fit to a halo  of only a few
hundred particles may carry large errors. We therefore make a rigorous
attempt to estimate  the errors and take  them  into account in  every
step  of the process.  Poor  fits due  to  small number statistics are
marked by  large errors that are  incorporated in the analysis and the
results we present.

The   profile   fit of haloes    in  crowded regions  clearly involves
ambiguities in the mass assignment to the subhaloes and the host.  Our
fitting procedure provides a  well-defined recipe for mass  assignment
based on the value of $\Mvir$ even  when the fit is actually performed
inside  an  $\rt$  that is   smaller  than  the  $\rvir$ obtained   by
extrapolation.\footnote{For 5\% of the subhaloes we actually find $\rt
<  \rs$. In these   cases the errors  in  the  extrapolated  values of
$\rvir$  and $\Mvir$ become   especially large, but they are  properly
taken into account in the  analysis.}  The concentration parameter  is
defined in the same way for all  haloes, $\cvir = \rvir/\rs$.  Because
in most cases of subhaloes  the extrapolation procedure adds much less
mass than the  mass that actually  lies between $\rt$ and $\rvir$, the
double counting is not  severe; most of the  mass associated with  the
upturn in the profile  is assigned to  a  different subhalo or to  the
host.  On the other hand, a small subhalo does not cause a significant
upturn in the profile of its host halo, and its mass is therefore also
included  in  the  mass  assigned to  the  host.   This partial double
counting introduces  some  uncertainty to  any recipe  for assigning a
luminous galaxy mass to a halo of a given mass.

The  outcome of  the halo  finder/classifier  is  a  statistical  halo
catalog that    includes  all the  bound    virialized systems  in the
simulation  above the   minimum mass  threshold. We include   distinct
haloes and subhaloes, but not subhaloes of  a second generation, i.e.,
those whose  hosts  are themselves  subhaloes of  a larger host.   The
output for each halo  includes the list of   its bound particles,  the
location of its centre, the NFW  profile parameters (e.g., $\cvir$ and
$\Mvir$), the  corresponding errors  ($\sigma_c$ and  $\sigma_M$), and
the truncation radius, if relevant.

\subsection{Tests of resolution} 
\label{sec:sims:tests}

\begin{figure}
\PSbox{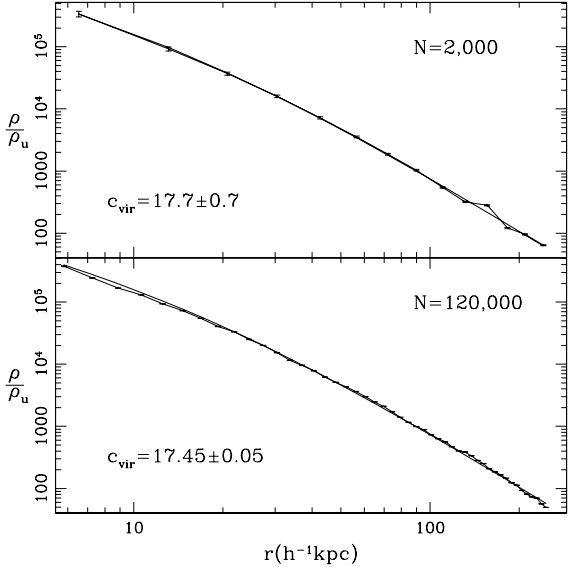 hoffset=-10 voffset=-65 hscale=40 vscale = 
40}{74mm}{74mm}
\caption{Convergence test for an $\Mvir = 2 \times 10^{12} \hMsun$ halo
simulated with $2,000$ and $120,000$ particles respectively.  When the
fit  is restricted to $0.02  \Rvir - \Rvir$  the best-fit $\cvir$
values show no significant difference.}
\label{fig:test1}
\end{figure}

\begin{figure}
\PSbox{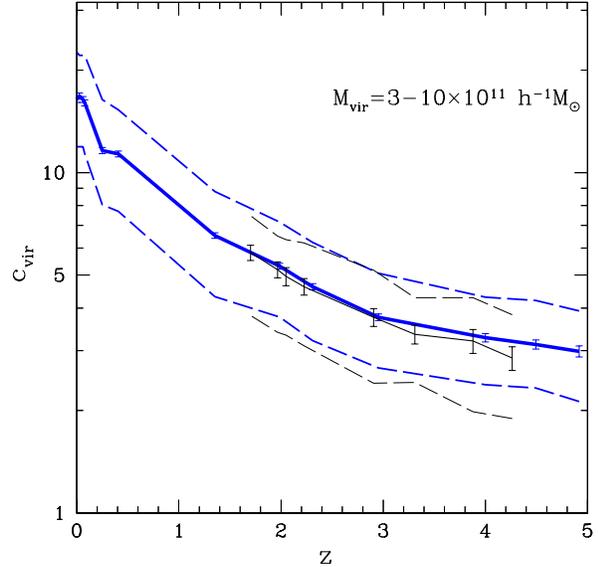 hoffset=-10 voffset=-65 hscale=40 vscale = 
40}{74mm}{74mm}
\caption{Convergence test for $\cvir$ evolution and scatter.
 Shown is  a  comparison  of 
$\Mvir = 3-10  \times   10^{11} \hMsun$ 
haloes simulated using our main simulation  (thick lines) and a
second simulation with $8$ times the mass resolution (thin lines).
The solid lines and errors reflect  the median and Poisson uncertainty
respectively.   The   dashed lines  reflect  the  estimated intrinsic
scatter.  There is no evidence for significant deviations in either
the measured median or scatter as the mass
resolution is increased.}
\label{fig:test2}
\end{figure}

In this  section we discuss the  effect of  varying mass resolution on
the measured fit parameters of haloes.  
Resolving the detailed shape of
halo density profiles at radii $r \lsim 0.01 \Rvir$ 
requires a sufficient
number of particles within the halo
(KKBP00; Moore et al. 1999), and one may worry that
our mass resolution will inhibit our ability to determine density
profiles for all but the most massive halos in our sample.  
However, a study of the inner slope of density profiles at small radii
is not the aim of this investigation.
We aim to  characterize  the general shape of  halo profiles
from  $0.02 \Rvir - \Rvir$ by measuring $\cvir$ values of halos, 
and our tests explore how mass resolution
affects results  when the fits are constrained to
this outer radial range.  

The first  test of mass resolution  effects was performed
using  a multiple mass  resolution version of the ART  code 
(KKBP00) in which a
halo  of interest is  simulated  with fine mass  resolution, while the
surrounding regions are   simulated   with particles of  masses   that 
are increasing as a function of distance  from the halo. The simulation of
a   particular   halo  was   performed   with  varying  maximum   mass
resolution. 
In Figure~\ref{fig:test1} we compare density  profiles of {\em the same} halo
of mass $1.5\times10^{12}\hMsun$ at $z=0$ in  the highest and the lowest
resolution  runs that  simulated  a box of  30$\hMpc$  on  a side. The
cosmology was    a  low-density  flat $\Lambda$CDM model     for which
$\omm=0.3$, $\oml=0.7$, $h=0.7$, and $\sigma_8=0.9$.

The high  mass  resolution  simulation was    run  with a peak   force
resolution of   0.2$\hkpc$  and  $500,000$ timesteps   at  the deepest
refinement level.  The  smallest mass particle in  this run has a mass
of $1.7\times10^7 \hMsun$, implying   an effective particle  number of
$120,000$ within  the virial radius of  the halo.  The  setup was such
that no  particles  other than those  of  the smallest  mass ended  up
within  the central $\sim 100  \hkpc$.   The lowest resolution run was
similar except  that the mass resolution  was such that the final halo
had 60 times fewer  ($2000$) particles within  its virial radius.   We
fitted  the resultant  halo profiles  in the  two runs  using the same
procedure outlined in the previous section and obtained $\cvir = 17.45
\pm 0.05$ and $\cvir = 17.70 \pm 0.70$ for the  highest and the lowest
mass  resolution runs,  respectively.  The  fits  are shown in the two
panels of Figure~\ref{fig:test1}.   The obtained values of $\cvir$ are
consistent  within   errors,  despite    the   vastly different   mass
resolutions.  This  gives us confidence in  our fitting  procedure and
resulting values of the concentration parameter.

In order to  extend our tests to  smaller particle numbers and to test
the dependence of the measured scatter and redshift behavior of fitted
$\cvir$ values (see \S 5.3 and 7), we have compared the results of the
$\Lambda$CDM simulation used for  our main analysis with a  simulation
with the  same cosmological parameters but  with half the box size and
$8$ times  the  mass  resolution.    This simulation was   stopped  at
$z=1.7$, so  we pursued  this  test only at  that  epoch  and earlier.
Because  the simulation of  higher resolution contains only few haloes
near  the high mass end,  we  limit the  comparison to  the mass range
(3-10)$\times 10^{11} \hMsun$.   The implied particle numbers for  the
low and high resolution simulations  are $300-1,000$ and $2,400-8,000$
respectively.  The evolution of measured $\cvir$  values as a function
of $z$ for  the  two simulations  is shown in  Figure~\ref{fig:test2}.
The solid lines are  the median relations, the  error bars reflect the
Poisson uncertainty  from the finite number of  haloes, and  the dashed
lines are the estimated intrinsic scatter (see \S 5.3 for a discussion
of our determination of  intrinsic scatter).  In  each case, the thick
lines correspond to the lower mass  resolution simulation and the thin
lines correspond  to the higher resolution  case.  Both the median and
the intrinsic spread   in $\cvir$ agree  to  within $\sim  5\%$.  This
level   of  agreement is consistent   with  the Poisson   error of the
high-resolution simulation, which is of order $10\%$ in the mass range
we consider.

We conclude that our mass resolution is perfectly adequate for the
purposes of this paper.

\section{HALO PROFILES TODAY}
\label{sec:today}

We start by  studying the halo profiles  at the  current epoch in  the
simulation.  First, we study  the median $\cvir$  (which is also close
to its  mean) as  a function  of  mass. Then,  the  dependence  on the
environment and  on being a subhalo is  presented.  Finally, we discuss
the scatter about the median $\cvir(\mu)$, which leads to a discussion
of   possible  implications on the  observed  rotation  curves and the
Tully-Fisher relation in the following section.

\subsection{Median relations for distinct haloes}

\begin{figure}
\PSbox{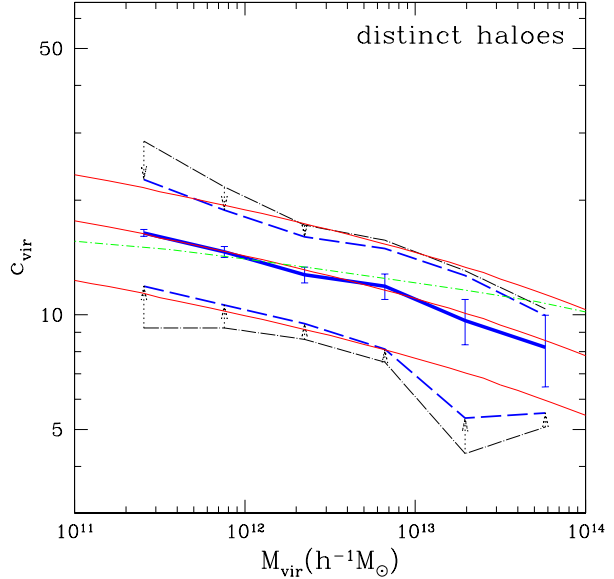 hoffset=-10 voffset=-65 hscale=40 vscale = 
40}{74mm}{74mm}
\caption{Concentration versus mass for distinct haloes at $z=0$.  
The thick solid  curve is the  median at a  given  $\mvir$.  The error
bars represent Poisson  errors of the mean  due  to the sampling  of a
finite number  of haloes per  mass bin.   The outer dot-dashed  curves
encompass $68\%$ of the $\cvir$ values as measured in the simulations.
The inner dashed curves represent only the  true, intrinsic scatter in
$\cvir$, after eliminating both  the Poisson  scatter and the  scatter
due to errors in the individual profile fits due,  for example, to the
finite number of particles per halo.  The central and outer thin solid
curves are the predictions for the median and $68\%$ values by the toy
model outlined in the text, for $F=0.01$ and three different values of
$K$.  The thin dot-dashed line  shows the prediction  of the toy model
of NFW97 for $f=0.01$ and $k=3.4\times10^3$.} 
\label{fig:c.vs.mass.distinct}
\end{figure}

Figure~\ref{fig:c.vs.mass.distinct}  shows $\cvir(\mvir)$ at $z=0$ for
distinct haloes.   The  Poisson errors  reflect the number   of haloes
within each mass  bin.   In order to account   for the fit errors,  we
generated 500 Monte Carlo  realizations in which the measured  $\cvir$
and $\mvir$ of each halo were perturbed using random Gaussian deviates
with   standard   deviations  equal   to   $\sigma_c$  and  $\sigma_M$
respectively.  Median values of $\cvir$ were then determined using the
Monte Carlo ensemble.  The lowest-mass and highest-mass bin have $\sim
2000$ and $20$ haloes respectively (we avoid using bins with fewer than
$10$ haloes.), and the Poisson errors grow with mass accordingly.

\begin{figure}
\PSbox{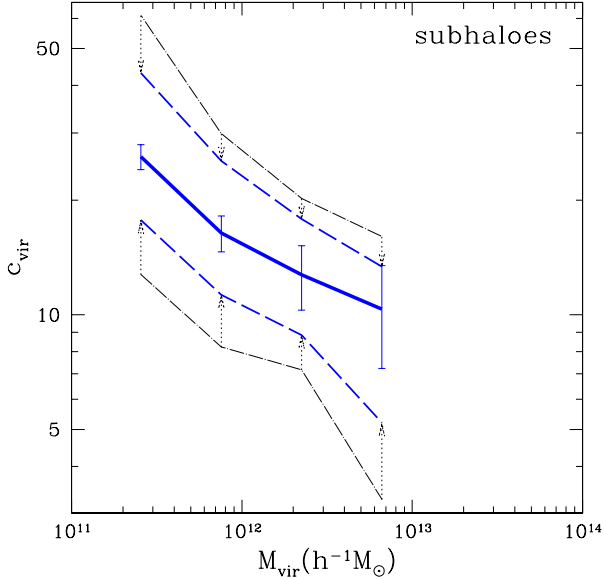 hoffset=-10 voffset=-65 hscale=40 vscale = 
40}{74mm}{74mm}
\caption{Concentration versus mass for subhaloes at $z=0$. The curves 
and errors are the same as in Figure~\ref{fig:c.vs.mass.distinct}.}
\label{fig:c.vs.mass.sub}
\end{figure}

The  median  $\cvir$  decreases   with growing  mass,   in qualitative
agreement  with the  toy  models, and   therefore consistent with  the
assertion  that small mass  haloes  are more concentrated because they
typically  collapse earlier than haloes  of larger  masses.
\footnote{The possibility that this trend is accentuated by
 the tendency for substructure to  be better resolved in
large-mass haloes has been investigated.  We  conclude that this is not
the case, since we  observe that the  number of subhaloes  within fixed
mass hosts does not correlate  with the measured concentrations.}   The
NFW97 model outlined in Appendix A has been slightly adjusted to yield
$\cvir$ rather  than $\cnfw$.  Its  predicted  slope is in  reasonable
agreement with that  derived from the simulations,  but  there is some
indication that the slope is too shallow.  Using our revised toy model
outlined in \S~\ref{sec:toy},  with $F=0.01$ and $K=4.0$, we reproduce
the median relation  even better, as shown by  the central  thin solid
line.  Near $\mu \sim 1$ ($\mvir \sim \Ms \simeq 1.5\times10^{13}\hMsun$
at $z=0$), the model prediction is
\beq
 \cvir(\mu,z=0) \simeq 1.25 K 
  \left(\mu F\right)^{- \alpha (\mu F)} \simeq 9 \mu^{-0.13}.
\label{eqt:cvm1}
\eeq
Indeed, the slope of the $\cvir(\mu)$  curve is closely related to the
varying   slope of  the  mass   power spectrum,  which  influences the
relative  difference in   collapse  epochs  for   typical objects   on
different   mass  scales.   The factor    of  $1.25$,  as explained in
\S~\ref{sec:toy},  is    a    measure  of  the   deviation    from the
Einstein-deSitter     self-similar linear  fluctuation growth     rate
$D(a)\propto   a$ between some  high   redshift and  $z=0$  (where the
corresponding  collapse epoch, for a   given mass,  is earlier in  the
$\Lambda$CDM case).

\subsection{Subhaloes and environmental dependence}

If  the median  $\cvir(\Mvir)$   indeed  reflects different  formation
epochs, one might expect the $\cvir$ of haloes of a given mass to vary
with  the density of  the environment, since  haloes  in dense regions
typically collapse  earlier.  In  particular, the concentration should
tend to be larger for subhaloes compared to  distinct haloes.  Another
effect that  may  lead to higher  $\cvir$  values in subhaloes is  the
expected   steepening     of  their  outer    profile   due  to  tidal
stripping~\cite{overmerge,ghigna,okamoto,aveela}.   Since stripping is
likely to be  more effective for small mass  haloes,  this process may
lead to a stronger  mass dependence in  subhaloes.  A third  effect is
that haloes that are embedded in a high-density environment are likely
to experience  extreme collapse  histories and frequent  merger events
which may affect their final concentrations.

Figure~\ref{fig:c.vs.mass.sub}   shows the relation $\cvir(\Mvir)$  at
$z=0$ for subhaloes.  We see that subhaloes on galactic scales ($\Mvir
\sim 10^{12}\hMsun$) are indeed more concentrated than distinct haloes
of the same mass.   The dependence  on  mass seems to be  stronger for
subhaloes  than for distinct   haloes, with $\cvir \propto \mu^{-0.3}$
(compared  to $\mu^{-0.13}$), though the  large errors in  the case of
subhaloes make this trend only marginal.

\begin{figure}
\PSbox{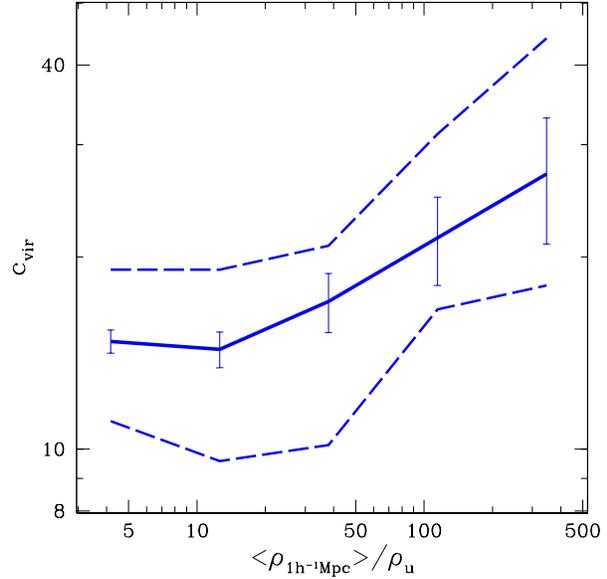 hoffset=-10 voffset=-65 hscale=40 vscale = 
40}{74mm}{74mm}
\caption{Concentrations versus environment.
The   concentration at  $z=0$   of  all  haloes in   the    mass range
$0.5-1.0\times10^{12}\hMsun$ as  a function of  local density in units
of  the average  density of  the    universe.  The local density   was
determined   within spheres  of    radius  $1\hMpc$.  The  solid  line
represents the median $\cvir$ value,  the error bars are Poisson based
on  the number  of  haloes, and  the  dashed  line indicates  our best
estimate of the intrinsic scatter.  }
\label{fig:c.vs.density}
\end{figure}

We address directly  the  dependence  of concentration  on  background
density in Figure  ~\ref{fig:c.vs.density},  which shows $\cvir$ as  a
function of local density for all haloes (both distinct and subhaloes)
in the mass  range $\Mvir=$0.5-1.0$\times 10^{12} \hMsun$.  The  local
background density  is defined for the  dark  matter within spheres of
radius $1\hMpc$ in units of the average density of the universe in the
simulation, $\rho_u = 8.3\times10^{10} h^2\Msun/$Mpc$^3$.  We see that
haloes in more dense environments  indeed tend to be more concentrated
than their  more isolated  counterparts.  Note  that this  trend is in
fact stronger than the dependence of $\cvir$ on mass.

We  find a similar trend  when the local  density is determined within
spheres  of  radius  $1.5\hMpc$,  but the trend   becomes  weaker when
spheres of radius  $0.5\hMpc$ are used.   Similarly, we find  that the
trend  holds for  haloes  of mass  $\lsim 5\times10^{12}\hMsun$.   For
larger masses, the trend seems to become  less pronounced, but this is
quite inconclusive  because we have only  a few massive haloes ($\lsim
15$) with   local densities $\gsim 100 \rho_u$.    A similar trend has
been seen for distinct  haloes alone, but  it  is difficult to make  a
definitive assessment  in  this case because   there  are only  a  few
distinct haloes with local densities $\gsim100\rho_u$.

The results  at  low and high densities   are comparable to  those for
distinct   haloes  and   subhaloes  respectively, consistent   with  a
significant correlation  between being a  subhalo (or a distinct halo)
and residing in a low-density (or a high-density) environment.

Our toy   model is not    sophisticated  enough  to account  for   the
dependence of concentration on environment, and for the exact relation
for   subhaloes.   However, in    the   framework  of  our toy   model
(\S~\ref{sec:toy}),  we can artificially  parameterize these trends by
varying the collapse parameter $K$ as a function of local density.  We
are left  with the qualitative  speculations  mentioned above for  the
interpretation of the trends with environment seen in the simulations.

We discuss possible implications   of  the environment  dependence  in
\S~\ref{sec:conc}.

\subsection {Scatter in the concentration parameter}

One of the most interesting of  our results at $z=0$  is the spread in
$\cvir$  values  for fixed  $\Mvir$.   A significant scatter about the
median relations  may   have  intriguing  observational  implications.
Before we report  on our results, we briefly  describe  our methods of
ascertaining the intrinsic scatter.

There are two  sources of scatter on top  of  the intrinsic spread  in
halo concentrations.  First, Poisson noise   due to the sampling of  a
finite number of  haloes in each mass bin  adds a  significant spread,
especially in the case of large-mass haloes. Small-mass haloes, on the
other hand,   are plentiful,  but   the  relatively  small  number  of
particles in each halo introduces  a significant error in the measured
value of the halo profile parameters $\Mvir$ and $\cvir$.  This is the
second  source of additional scatter.   The  Poisson error due to  the
finite number of haloes is relatively  straightforward to correct, but
correcting  the error  in   the  profile  parameters requires  a  more
involved procedure.

We account  for this error in  the profile parameters using the errors
obtained in the profile fits.  Within each mass bin, we have performed
$\sim500$ Monte Carlo realizations in an attempt to undo the effect of
the  measurement errors as follows.  Every  measured $\cvir$ value has
been perturbed   by a one-sided  random Gaussian  deviate, positive or
negative depending  on whether   the  measured $\cvir$ is  smaller  or
larger than the  median respectively.  The  standard deviation of each
Gaussian  deviate was set  to be the error in  the value of $\cvir$ as
estimated in the  profile  fit of   that specific  halo.   The smaller
scatter obtained  in this set of  Monte Carlo realizations provides an
estimate for the spread excluding the fit errors.  We then subtract in
quadrature the Poisson error  due to  the  finite number of  haloes to
obtain our estimate for the intrinsic scatter in $\cvir$.

We have checked our technique of  measuring the intrinsic spread using
an  artificial ensemble of 1000  (spherical)  haloes with a variety of
numbers of particles and   a known intrinsic distribution of  $\cvir$.
The technique  reproduced the median concentration  and true spread to
within $5\%$ when the particle number was varied from $100$ to $10^5$,
the range of interest for our simulated haloes.

As discussed in \S 4.3, we have also checked 
our procedure for measuring the intrinsic scatter
using a  simulation of  higher resolution  in  a  smaller box  of side
$30\hmpc$, in which  there are on average $8$  times as many particles
in a halo of a given mass.  Although the measured scatter
was larger in the lower resolution simulation, the estimated
intrinsic intrinsic  spreads in the two simulations 
agree to   within $\sim  5\%$ (see Figure~\ref{fig:test2}).  This
gives us confidence in our technique.

Note that  what we  treat  as measurement  error  in the  profile  fit
actually includes scatter due to real deviations of the halo structure
from a purely  spherical NFW profile,  which we should probably regard
as part   of  the intrinsic  scatter. This  means  that  our estimated
intrinsic scatter is a conservative underestimate.

\begin{figure}
\PSbox{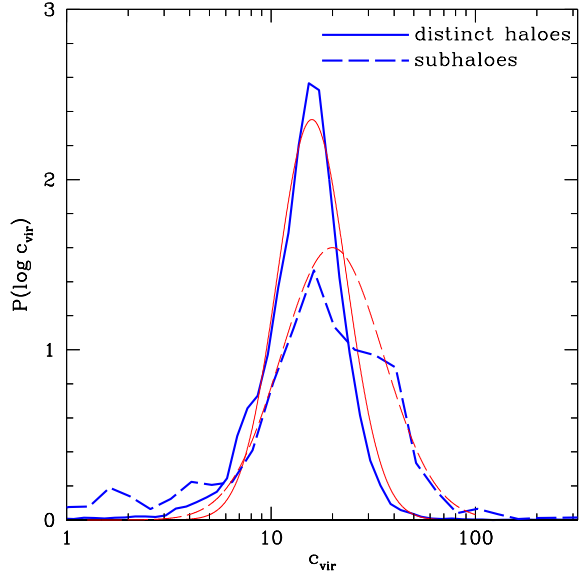 hoffset=-10 voffset=-65 hscale=40 vscale = 
40}{74mm}{74mm}
\caption{The probability distributions of distinct haloes (solid line) 
and subhaloes (dashed    line)   at $z=0$   within  the   mass   range
$(0.5-1.0)\times10^{12}\hMsun$.   The  simulated  distributions (thick
lines)  include,  the $\sim 2,000$   distinct   haloes and $\sim  200$
subhaloes  within this mass range.   Log-normal distributions with the
same  median and standard  deviation as the measured distributions are
shown (thin lines).  Subhaloes are, on average, more concentrated than
distinct haloes and they show a larger spread.  }
\label{fig:subpardist}
\end{figure}


Now that the method has been discussed, we  turn the attention back to
the    relation  between $\cvir$   and  $\Mvir$  for  distinct haloes,
Figure~\ref{fig:c.vs.mass.distinct}.  The  measured $68\%$  scatter is
shown, as well as  the ``pushed in" corrected  scatter which marks our
(under-) estimated intrinsic  scatter.  As can  be seen by noting  the
Poisson error bars,  the correction at   the small-mass end  is almost
entirely  due  to the measurement errors    of the profile parameters,
which are dominated by the small number of particles per halo.

We see that  the intrinsic spread  is large;  it is  comparable to the
systematic change in  the median  value of  $\cvir$  across the entire
mass range studied.  In addition, the spread  is roughly constant as a
function of mass, with a $1\sigma$ deviation of $\Delta(\log\cvir)\sim
0.18$.  We discuss possible observational implications of this scatter
in the next section.

The spread in $\cvir$ values as a function of $\Mvir$ for subhaloes is
shown in  Figure~\ref{fig:c.vs.mass.sub}.   Note that  the scatter  is
larger for the subhalo population than for their distinct counterparts
of   the same mass, with  a  $1\sigma$ variation of $\Delta(\log\cvir)
\sim 0.24$.   This is  clearly  seen in   Figure~\ref{fig:subpardist},
where the  probability  distributions of  concentrations for  distinct
haloes and subhaloes are compared  (for $\Mvir = 0.5-1.0 \times10^{11}
\hMsun$).    It  is possible  that  the larger   scatter evaluated for
subhaloes is  a result of  their more complicated formation histories,
including for  example more interactions  and stripping.  We point out
that  we  have   found no   significant  trend   with the   number  of
co-subhaloes within the same virialized host.  Such a trend might have
been expected if interaction  among  co-subhaloes plays an   important
role in determining the profile shape.

The $68\%$ intrinsic   spread in $\cvir$  as  a function of  the local
density   (for  $0.5-1.0\times10^{12}\hMsun$  haloes)    is  given  in
Figure~\ref{fig:c.vs.density}.  We  can  use the obtained distribution
of   halo concentrations  as  a  function of  local   density to probe
questions associated with the origin  of LSB galaxies and the observed
morphology density relation (\S~\ref{sec:conc}).

In   Figure~\ref{fig:subpardist}    we   show    the  distribution  of
concentration values for  distinct haloes  and subhaloes, along   with
log-normal functions with the same median and standard deviation.  The
log-normal forms describe the  observed distributions reasonably well.
Such a result  has also  been  reported by Jing (2000).   Our measured
scatter  for distinct haloes is similar  to that  reported by Jing for
equilibrium haloes.

In the context of the toy model presented in \S~\ref{sec:toy}, one can
parameterize the spread in $\cvir$ as spread in collapse epochs and/or
collapse  histories,  via the  parameters $\ac$  and $K$ respectively.
Using Eq.~\ref{eqt:toy4}, we find that the evaluated spread in $\cvir$
can  be  matched by    a  spread of   $\Delta[\log K(a_0/\ac)]  \simeq
\Delta(\log\cvir) \sim 0.18$ in the toy model.
If we absorb  all the scatter  in the collapse  parameter $K$, we find
that  the   model  matches  the   $50\pm34\%$  (encompassing   $68\%$)
percentiles  of  the $\cvir$   distribution  with  $K=6.0$  and  $2.6$
respectively  (for $F=0.01$).   Note that the  scatter in   $K$ is not
symmetric  about the  median  (of  $K=4.0$);  it  rather reflects  the
log-normal nature of the  $\cvir$ distribution.  The model predictions
for  the above  values   of $K$  are  shown as   thin  solid lines  in
Figure~\ref{fig:c.vs.mass.distinct}; they match the simulated  scatter
fairly   well.    We   show   in   \S~\ref{sec:redshift}    that  this
parameterization also reproduces the simulated  scatter as a  function
of $z$.
This is just a  useful parameterization of  the scatter using  the toy
model.  A  more detailed modelling  of the spread with deeper physical
insight is beyond the scope of the present paper.

\begin{figure}
\PSbox{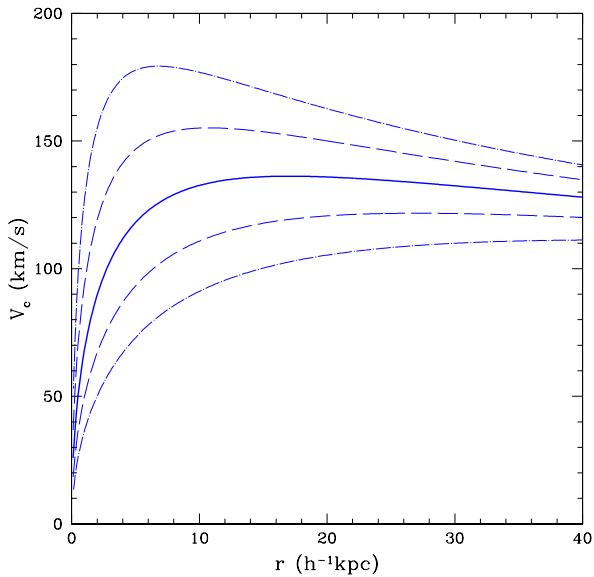 hoffset=-10 voffset=-65 hscale=40 vscale = 
40}{74mm}{74mm}
\caption{The spread in NFW rotation curves corresponding to the spread
in    concentration     parameters      for distinct       haloes   of
$3\times10^{11}\hMsun$  at $z=0$. Shown  are the  median (solid), $\pm
1\sigma$  (long dashed), and  $\pm 2\sigma$  (dot-dashed) curves.  The
corresponding median rotation curve for subhaloes is comparable to the
upper $1\sigma$ curve of distinct haloes.  }
\label{fig:vspread}
\end{figure}

\section{ROTATION CURVES AND TULLY-FISHER}
\label{sec:TF}

The simulated distributions  of $\cvir$ values  as a function of  mass
and  environment have   several observational  implications.  Here, we
discuss only preliminary predictions involving rotation curves and the
Tully-Fisher    relation  based on    very  crude   assumptions  about
associating disc  properties  to those  of the  simulated  dark-matter
haloes.   A  more  detailed study    requires realistic  modelling  of
physical processes involving gas and stars.

In order to  illustrate  what the spread  in $\cvir$  values may imply
observationally,  Figure~\ref{fig:vspread} shows example NFW  rotation
curves for $3\times10^{11}\hMsun$ (distinct)  haloes using the median,
$\pm1\sigma$,   and $\pm2\sigma$ values   of  $\cvir$.  These  are raw
rotation curves of the dark-matter haloes before  they are affected by
the infall  of  baryons,  but  they  may   still  serve as  a    crude
approximation for the final  rotation curves.   One  can see that  the
rotation   curves   span a significant    range   of  shapes and   the
corresponding spread in   $\Vmax$ values is substantial.   The  median
rotation curve  for $3\times10^{11}\hMsun$  subhaloes (not  shown)  is
similar   to     the       upper  $1\sigma$      curve    shown     in
Figure~\ref{fig:vspread}.
  
\begin{figure}
\PSbox{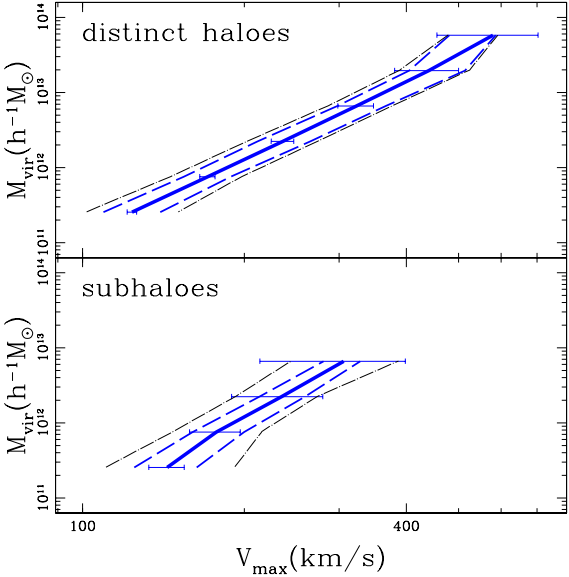 hoffset=-10 voffset=-65 hscale=40 vscale = 
40}{74mm}{74mm}
\caption{Virial mass versus $\Vmax$ at $z=0$ for distinct haloes (a) 
and subhaloes (b).  The   outer   dot-dashed and dark   dashed   lines
indicate  the    measured  and corrected   intrinsic    $68\%$ scatter
respectively.  }
\label{fig:MvsVmax}
\end{figure}

A clue for  the expected TF relation  of discs may  be provided by the
measured  relation between  the halo parameters   $\mvir$ and $\Vmax$.
The latter  is  derived from $\cvir$   using  Eq.~\ref{eqt:vmax}.  The
$\mvir$-$\Vmax$   relation  is  shown    in  Figure~\ref{fig:MvsVmax},
separately    for    distinct  haloes  and   subhaloes.     The median
distinct-halo relation is well approximated by the linear relation
\beq
\log[\Mvir/(\hMsun)] = \alpha \log[\Vmax/({\rm km/s})] + \beta
\label{eqt:Vlinear}
\eeq
with $\alpha \simeq 3.4 \pm 0.05$ and $\beta \simeq 4.3\pm 0.2$, where
the Poisson errors in each mass bin have been propagated to obtain the
quoted error on each fit value.   Note that the  slope is steeper than
that expected from   the standard scaling   of the virial  parameters:
$\Mvir \propto  \Vvir^3$.  This is a direct  result of the correlation
between mass  and concentration  ($\Mvir \propto \cvir^{-0.13}$).   We
may in fact derive the expected $\alpha$ using the effective power law
from   Eq.\ref{eqt:vmax}:  $\Vmax/\Vvir \propto  \cvir^{0.27}$;  which
implies $\Mvir \propto \Vmax^3 (\Vvir/\Vmax)^3 \propto
\Vmax^3 \cvir^{-0.81} \propto \Vmax^{3.4}$.   
We point out that the linear relation provides a good fit, showing no obvious
need for non-linear corrections in the TF relation. 

The lower panel in  Figure~\ref{fig:MvsVmax} shows $\mvir$ vs. $\Vmax$
for subhaloes.  This relation is also well fit by the linear relation,
Eq.~\ref{eqt:Vlinear}, but now  with $\alpha \simeq  3.9 \pm 0.25$ and
$\beta \simeq 2.6 \pm 0.75$.  The subhalo relation has a steeper slope
compared   to distinct  haloes,   and   $\Mvir \simeq   10^{12}\hMsun$
subhaloes  typically have    a $\sim  12\%$    higher  $\Vmax$.   This
difference  between the slope  and zero-point  of distinct haloes  and
subhaloes may  have implications  for  the  use of  cluster or   group
galaxies to calibrate the Tully-Fisher relation in the field.

We point out, however, that if for subhaloes we replace $\mvir$ by the
mass $\mt$ inside the  truncation radius $\rt$, the  logarithmic slope
becomes $\alpha_t=3.6\pm0.2$,  consistent with  the slope  obtained by
Avila-Reese et al. (1999)  for haloes within  clusters using a similar
mass assignment  procedure.  (The reason  for the slope change is that
the  ratio of $\mt/\Mvir$   is  roughly 1 for  low-mass,  high-$\cvir$
haloes, and becomes less  than  1 for high-mass, low-$\cvir$  haloes.)
This slope is similar to the slope we find for distinct haloes.  It is
not obvious a priori which of the halo  masses is more relevant to the
mass of the cooled gas that ends up as the luminous  disc, and thus to
the  observed Tully-Fisher  relation. Therefore,  the  worry about the
universality of the slope  of Tully-Fisher is not conclusive. However,
the  zero-point difference  between  the  two  types of haloes  exists
regardless of the mass choice, and is a robust result.

The scatter in the TF relation is an issue of great interest.  We find
for distinct haloes  a $1\sigma$ scatter at  fixed  $\Vmax$ of 
$\Delta (\log \Mvir)  \simeq 0.17$, 
while the corresponding scatter for   fixed $\Mvir$ is $\Delta   \Vmax
/\Vmax \simeq  0.12$.  This  scatter  is in rough  agreement with  the
spread predicted by Eisenstein \& Loeb (1996)  for a similar cosmology
using  Monte Carlo realizations of  halo formation  histories based on
the Press-Schechter approximation.  The subhalo relation shows an even
larger  scatter, with  $\Delta \Vmax /\Vmax   \simeq 0.16$ at  a fixed
$\Mvir$.

Observational estimates for the intrinsic scatter in (I-band) TF range
from $\sigma(V)/V \sim 0.09$ \cite{willick:96} to $\sim 0.03$
\cite{bernstein:94}.  At best, the observed scatter leaves no room for
any  intrinsic variation in  the mass-to-light  ratio of galaxies, and
may imply that gas contraction and other hydrodynamical processes must
somehow act to  decrease the scatter.   A simple idea that may resolve
this discrepancy is discussed in Appendix C and we briefly outline the
argument  here.  For a fixed mass  and  spin, a more concentrated halo
(higher $\vmax$)   will  induce more gas contraction,    and therefore
produce  a  smaller, brighter  disc.  Such a  correlation  between the
mass-to-light ratio of galaxies and  the deviation of $\vmax$ from the
median $\vmax$ at  a given $\Mvir$ could  reduce  the scatter to  that
required   to match observations.   Detailed  modelling, including the
back-reaction  of the  halo during  disc formation, is  needed to test
this hypothesis in detail.

\section{REDSHIFT DEPENDENCE} 
\label{sec:redshift}

As data accumulate at high redshift, it becomes increasingly important
to study the predicted evolution of the population of halo profiles as
a function of redshift.

\begin{figure}
\PSbox{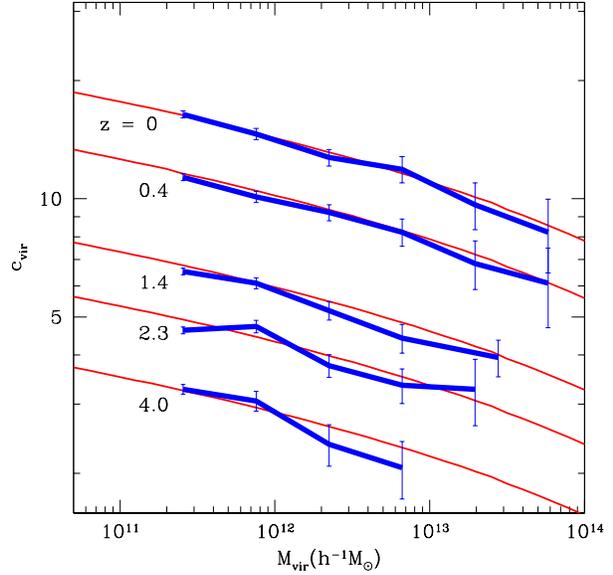 hoffset=-10 voffset=-65 hscale=40 vscale = 
40}{74mm}{74mm}
\caption{Median $\cvir$ values as a function of $\Mvir$ for
distinct haloes at various redshifts.  The  error bars are the Poisson
errors due to the finite number of haloes  in each mass bin.  The thin
solid lines show our toy model predictions.  }
\label{fig:cmz}
\end{figure}

Figure~\ref{fig:cmz} shows the median $\cvir$ as a function of $\Mvir$
for distinct haloes at several different redshifts.  We see that for a
fixed mass,   the typical  $\cvir$ value changes   quite dramatically,
while the shape of the mass dependence  remains roughly constant.  The
thin solid lines show  our toy model predictions.  This  two-parameter
model, which has been normalized  to match the slope and normalization
of the  relation at $z=0$, does remarkably  well at all redshifts.  As
predicted by the toy  model in \S~\ref{sec:toy}, the  concentration of
haloes of a fixed mass scales as $\cvir \propto (1+z)^{-1}$.
A similar behavior has  been confirmed using the  SCDM simulation in a
smaller box ($7.5\hMpc$, described in \S~\ref{sec:sims}).
The redshift dependence of the  subhalo concentrations seems  similar,
but  we   don't  have sufficient  statistics   for  conclusive results
involving subhaloes at high redshifts.

As    mentioned in the Introduction,    the  dramatic evolution in the
concentration of haloes    of a  fixed  mass  is   different from  the
prediction of  the NFW97 analytic toy  model (see Appendix A). This is
illustrated in Figure~\ref{fig:c.vs.z}, which shows the median $\cvir$
of  the distinct  halo   population  of  $\Mvir  =  (0.5-1.0)   \times
10^{12}\hMsun$ as  a  function of redshift.   The  NFW prediction (for
$0.8\times  10^{12} \hMsun$  haloes)  overestimates  $\cvir$ by  $\sim
50\%$ at $z=1$,  and the disagreement grows with  redshift.~\footnote{
As  discussed in Appendix  A, if the NFW model  is modified by setting
the free parameter $f$ to the unphysically small value $f = 10^{-15}$,
then  the redshift behavior can be   reconciled with what is observed.
However, for this  small  value of $f$,    the NFW model   predicts an
excessively  flat $\cvir$ vs.  $\Mvir$  trend.}  
Since NFW's approach was to resimulate a small number of halos identified 
at $z=0$ in a larger simulation, they could not check the redshift 
dependence of the halo concentration.  Their resimulations typically had 
not yet collapsed to a single halo at high redshift.
Our revised toy model
reproduces the simulated redshift trend very  well.  The scatter about
the relation is remarkably constant as a function of $z$: $\Delta(\log
\cvir) \sim 0.18$.  Also shown is  how the spread can be parameterized
by varying $K$ in our toy model, as discussed in
\S~\ref{sec:today}.

The redshift  dependence of the   inner radius, $\rs$, can  be deduced
from that of $\cvir$ by recalling that the virial radius of fixed-mass
haloes   also varies  like  $\rvir   \propto \Dvir^{1/3}/(1+z)$.  This
implies that, on average, the  inner radius of  haloes of a given mass
remains roughly constant as a function of redshift (aside from the $z$
dependence   of     $\Dvir$).     We   see   this    explicitly     in
Figure~\ref{fig:rs.vs.z.distinct},  which shows  the evolution of  the
median and $68\%$ scatter  of $\rs$ as a function  of $z$ for distinct
haloes in the mass range $0.5-1.0\times10^{12} \hMsun$.  The fact that
the median $\rs$ value declines slightly near $z=0$ is  due to the $z$
dependence    of $\Dvir$ in   the   $\Lambda$CDM model simulated.  The
robustness of the  characteristic length scale,  $\rs$, may provide an
interesting  clue for the   understanding of the  build-up  of DM halo
structure.

The strong decline in the concentration of haloes of a fixed mass as a
function  of   redshift    should  have  an    interesting  impact  on
galaxy-formation modelling    at high  redshift  ---  e.g.,   aimed at
understanding    the         nature            of     Lyman      Break
Galaxies~\cite{steidel:96,lowenthal}   and  the     evolution   of the
Tully-Fisher relation~\cite{vogt}.  Although,  in general, haloes, and
therefore  galaxies, are expected   to  be  smaller at  high  redshift
(reflecting the higher universal density) and  to have higher circular
velocities  ($\Vvir  \propto   \Rvir^{-1/2}$),  the observed   $\cvir$
behavior will tend to counteract this tendency.

Insight  into the  expected TF evolution  of discs  may be gained from
Figure~\ref{fig:TFevo},  which   shows  the  $\Mvir$   versus  $\Vmax$
relation  for  (distinct)  haloes   at several  redshift  steps.   The
evolution in  the  zero-point is indeed   less dramatic than  would be
expected from the  scaling of $\Vvir$.   In fact (not  shown) there is
almost no evolution in the zero-point between $z=0$ 
and $0.5$.  The slope of
the relation is roughly constant as  a function of redshift 
($\alpha = 3.4 \pm 0.1$) and the scatter is roughly constant;
$\Delta V_{\rm max}/V_{\rm max} \simeq 0.12$.

Furthermore, because disc size is expected to be a decreasing function
of halo concentration (\S 2), the decline  of $\cvir$ with $z$ implies
a relative  increase  in disc  sizes at  high  redshift.  This  should
result  in lower than expected  surface brightnesses at high $z$, both
because of the extended size and the corresponding lower efficiency of
star  formation.  To this one  could add the fact   that the supply of
cold  gas for disc  formation at  high  redshifts may be limited  (not
extending all the way to  $\rvir$) because the smaller inner densities
will  lessen  collisional   cooling.  These results    may  hinder the
association  of quiescently   star-forming  objects  with  Lyman-break
galaxies   as  discussed  e.g., by  Mo,    Mao, \&  White (1999)  (see
\citeNP{nature}; \citeNP{spf:99} for an alternative physical model for
Lyman-break galaxies).   They further argue for  the slow evolution of
the Tully-Fisher relation.

\begin{figure}
\PSbox{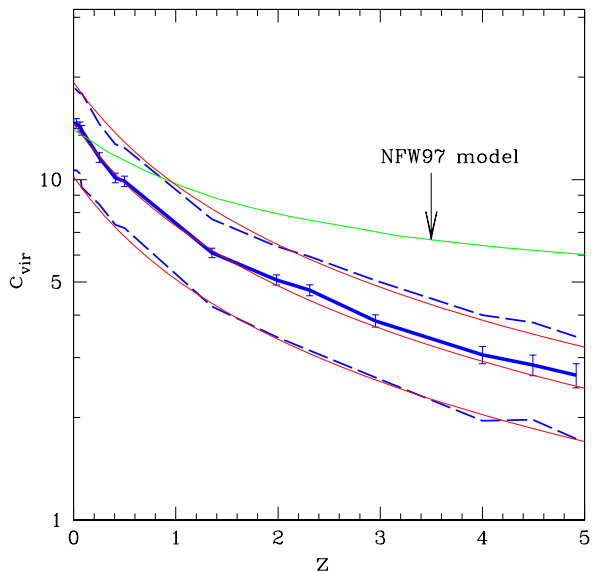 hoffset=-10 voffset=-65 hscale=40 vscale = 
40}{74mm}{74mm}
\caption{Concentration as a function of redshift for distinct haloes
of   a fixed  mass,  $\mvir=0.5-1.0\times10^{12}\hMsun$.    The median
(heavy solid   line) and  intrinsic  $68\%$ spread  (dashed  line) are
shown.  The behavior predicted by the NFW97 toy model  is marked.  Our
revised toy model   for the median  and spread  for $8  \times 10^{11}
\hMsun$ haloes   (thin solid lines)   reproduces the observed behavior
rather well.  }
\label{fig:c.vs.z}
\end{figure}

\section{CONCLUSIONS AND DISCUSSION}
\label{sec:conc}

The main  direct conclusions   of  this paper,  based on  analyzing  a
statistical sample    of dark-matter   haloes  in a    high-resolution
simulation of the $\Lambda$CDM cosmology, are as follows:
\begin{itemize}
\item 
The redshift  dependence  of the   halo  profile parameters  has  been
measured in the simulations, and reproduced  by an improved toy model.
For example, $\cvir \propto  (1+z)^{-1}$ for haloes  of the same mass,
predicting that at high  redshift they are  less concentrated and with
larger   inner  radii   than  previously expected.  The  corresponding
prediction  for  rotation curves  is lower values  of $\vmax/\vvir$ at
high $z$.
\item 
The  correlation  between any   two halo    profile parameters  has  a
significant scatter. For example, in the $\cvir$-$\mvir$ relation, the
spread in $\cvir$  is comparable to the  systematic  change in $\cvir$
across three orders of magnitude in $\Mvir$.  The $1\sigma$ spread for
fixed  $\Mvir$ is   $\Delta(\log\cvir) \simeq  0.18$, corresponding to
$\Delta \vmax/\vmax \simeq 0.12$ at a given $\mvir$.
\item 
There are   indications  for environmental trends  in   halo profiles.
Haloes in dense environments, or subhaloes, are more concentrated than
their isolated counterparts of the  same virial mass, and they exhibit
a larger scatter in $\cvir$.
\end{itemize}

The main implications of the above findings can be summarized as follows:
\begin{itemize}
\item
Disc galaxies at high redshifts are predicted  to be more extended and
of lower surface  brightness than  expected previously.  The  constant
inner  radius at fixed mass may  be a dynamical clue for understanding
the formation of halo structure.
\item
The scatter in the halo mass-velocity relation is significantly larger
than in the  observed TF relation,  which suggests that the luminosity
of a disc forming inside a halo of a given  mass should correlate with
the  maximum   rotation velocity.  We  pointed   out a possible simple
explanation for that.
\item 
The environmental trends  of   halo profiles may caution   against the
universality of the TF  relation.  In addition, these trends, together
with the observed scatter, may provide insight  into the origin of the
Hubble sequence.  Below, we   argue that haloes of  low  concentration
will tend to host blue galaxies  and haloes of high concentration, red
galaxies or spheroids.  We also  point out that extremely low  $\cvir$
haloes plausibly host LSB galaxies.

\end{itemize}

\begin{figure}
\PSbox{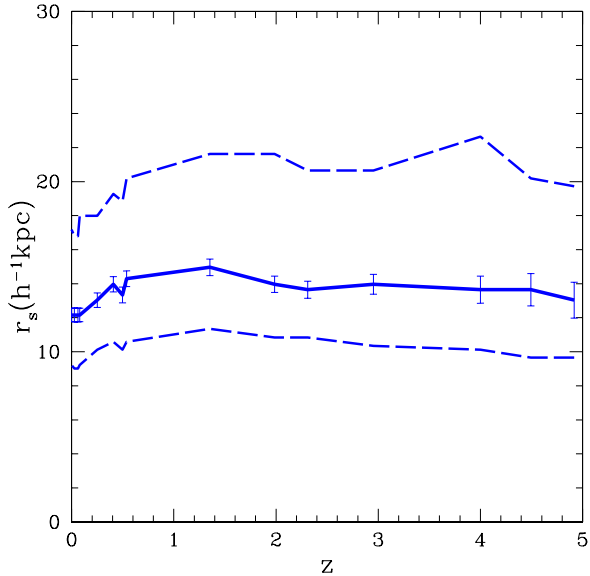 hoffset=-10 voffset=-65 hscale=40 vscale = 
40}{74mm}{74mm}
\caption{The inner radius $\rs$ as a function of redshift for distinct
haloes of  fixed mass, $\mvir=0.5-1.0\times10^{12}\hMsun$.   Shown are
the  median (solid line)  and  intrinsic $68\%$ spread (dashed lines).
The    median value for  $\rs$    remains approximately constant  as a
function of redshift.  }
\label{fig:rs.vs.z.distinct}
\end{figure}

\begin{figure}
\PSbox{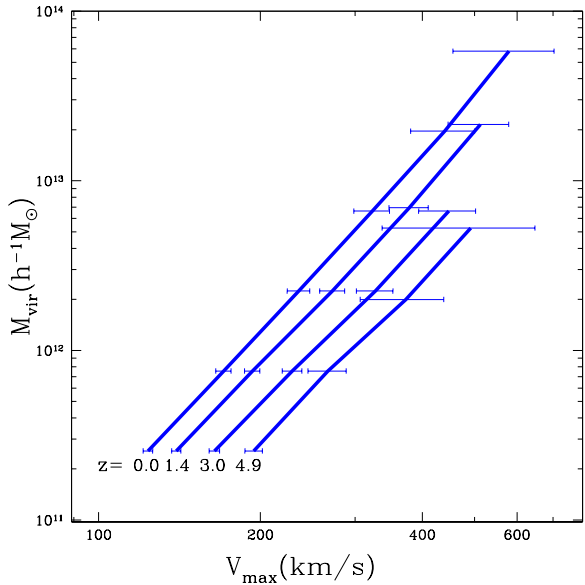 hoffset=-10 voffset=-65 hscale=40 vscale = 
40}{74mm}{74mm}
\caption{The evolution of the distinct halo Tully-Fisher 
relation, $\Mvir$  versus $\Vmax$,  for  several redshift steps.   The
evolution  is weaker than the $\Mvir$  versus $\Vvir$ relation because
$\cvir$ falls rapidly with redshift (Fig.~\ref{fig:c.vs.z}).  }
\label{fig:TFevo}
\end{figure}

We have  proposed an alternative  to the toy model originally proposed
by NFW96.   It reproduces the  correlations between the two parameters
of   the halo profiles,   e.g,  $\cvir$ and  $\mvir$,  as  well as the
redshift dependence  of these correlations.  This  model also offers a
simple parameterization  that reproduces the  scatter about the median
relation  observed  in our simulation.  The   modified toy model  is a
useful tool for   semi-analytic modelling  of   galaxy formation.   In
particular,   analyses of the   type performed by~\citeNP{MMW2}, which
made predictions  for disc properties at  $z\sim 3$, based on the halo
toy  model  of NFW97, should be   reconsidered using our  modified toy
model.   A program that implements the model for several cosmologies 
and provides $\cvir(\Mvir,z)$ with the  expected
scatter is available from the authors upon request.

The large intrinsic  scatter we find  in the correlations  between the
halo  profile parameters makes the  haloes  a two-parameter family, as
expected,  and should be taken into  account  when trying to model the
scatter in  observable properties of galaxies  (c.f. Mo, Mao, \& White
1998; Navarro 1998; Navarro \& Steinmetz 1999).

The spread in exponential disc sizes implied from our $68\%$ spread in
concentration  values  for  an  $\Rvir  =   200 \hkpc$   halo at $z=0$
($\Mvir\,\sim\,  10^{12} \hMsun$, $\cvir$:$9.2  \rightarrow 21.3$)  is
$\rd:4.0 \rightarrow 2.6\hkpc$   (see 
Eq.~\ref{eqt:rd}).  This   is
roughly the same spread in disc sizes  resulting from a spin parameter
variation of  $\lambda:0.05\rightarrow 0.03$, which  is  approximately
the intrinsic  spread  in $\lambda$ inferred  from  N-body simulations
(\citeNP{be:87,warren,spin}).

The  two quantities,  $\cvir$ and $\lambda$,   are thus of  comparable
importance for determining   observable properties  of  galaxies,  and
define a plane in   parameter  space for haloes   of fixed  mass.   We
suggest that the $\cvir$-$\lambda$ plane can  perhaps be linked to the
observed variations of galaxies and help  explain the Hubble sequence.
For example, we argued above that  haloes with very low concentrations
would tend to lead to discs of low surface brightness, and with slowly
rising rotation curves  (i.e.,  $\vmax/\vvir \sim 1$).   This argument
will only apply,  however,  if $\lambda$ is  large  enough to  prevent
excessive gas infall  to the center of the  halo, which would tend to
increase the effective concentration of the system.  Similarly, haloes
of  high-$\cvir$ and low-$\lambda$ will   likely be unable to  produce
dynamically stable discs (see, e.g., Mo et al. 1998), and instead host
spheroids.  Other combinations of these parameters, may, perhaps, lead
naturally to a range of galaxy morphologies.

This  kind of mapping is further  motivated by  the understanding that
high-$\cvir$ haloes  collapse  earlier   than low-$\cvir$  haloes  (as
predicted by the toy model and explicitly  demonstrated by NFW97 using
simulated haloes).  A natural  association  is then that  high-$\cvir$
haloes host  old, red galaxies, and  lower-$\cvir$ haloes  host young,
blue galaxies.  Furthermore,  the environmental trend, that haloes  in
low-density environments  tend to be  less concentrated than haloes of
the same  mass in  high-density environments,  fits nicely   into this
picture.  Indeed,  LSB galaxies are observed to  be more isolated than
galaxies of higher surface brightness
\cite{bothun:93,mo:94}, and spheroids
tend to inhabit high-density environments
\cite{dressler:80,postman:84}.

We have made  in this paper only crude  preliminary  attempts to study
the implications of our results concerning halo profiles.  
In Bullock et al. (2000) we have performed a first step towards understanding
how $\cvir$ evolution affects the nature of the
galaxy luminosity-velocity relation at high redshift.
It would be desirable  to  go on and   
incorporate  our    measured  halo   properties, including scatter,  into
semi-analytic modelling of gas  processes and star formation in  order
to  make  detailed predictions  for observable  galaxy  properties and
their evolution.  

\section*{Acknowledgments} 
We would like to thank Andi Burkert, Alberto Conti, Marc Davis, Daniel
Eisenstein,  Sandy Faber, Avi Loeb, Ari   Maller, Julio Navarro, Jason
Prochaska, David  Spergel, Frank van den  Bosch, Risa  Wechsler, David
Weinberg, and  Simon White for helpful  discussions  and comments.  We
thank the anonymous referee  for useful comments and suggestions.  The
simulations were  performed at NRL and  NCSA.  This work was supported
by  grants from NASA and  NSF at UCSC and NMSU,  and by Israel Science
Foundation and US-Israel Binational Science Foundation grants.

\appendix
\section{The NFW97 model}

For   completeness,    we briefly  review the   NFW   prescription for
determining $\cvir(M,a)$.  The goal of the NFW procedure is to provide
the density  profile of a  halo of mass  $M_{\rm vir}$  at epoch $a_0$
assuming that the  profile is  of  the NFW form (\ref{eqt:nfw}).   The
collapse  redshift    is now  determined using     the Press-Schechter
approximation,  which, given  $\Mvir$ at  epoch $a_0$, can  be used to
approximate the probability distribution for the epoch $a$ when a halo
trajectory was  first more massive  than some fraction $f$  of $\Mvir$
(Lacey \& Cole 93)
\beq
P(>f \Mvir,a|\Mvir,a_0) =
{\rm erfc}\left[\frac{\delta_{\rm crit}(a) - \delta_{crit}(1)}
     {\sqrt{2(\sigma_0^2(f \Mvir) - \sigma_0^2(\Mvir))}} \right].
\label{eqt:nfw1}
\eeq
The quantities $\sigma_0$   and $\delta_{\rm  crit}$ are  defined   in
\S~\ref{sec:toy}.      One      determines   $a_{c}$,     by   solving
Eq.~\ref{eqt:nfw1} for the
\textit{most probable} value of $a$ by setting
$P(>f\,M,\ac|M,a_0) = 0.5$.  One  now assumes that the central density
of the halo is proportional to the density of the universe at $a_{c}$,
which implies:
\beq
        \rho_s = k \rho_{u}(a_0)\left[\frac{\ac}{a_0}\right]^3,
\label{eqt:nfw2}
\eeq
where $k$   is a numerical constant.    Now, given $\Mvir$  and $a_0$,
Eqs.~\ref{eqt:nfw1} and  \ref{eqt:nfw2}   determine $\rho_s$  and thus
completely specify the density profile.

This  procedure has two  free  parameters, $f$ and  $k$,  which may be
adjusted in order to match the slope and normalization respectively of
$\cvir (M)$  at $a=1$.   NFW97   show that  this  two  parameter model
reproduces  the $a=1$   relations  of  simulated  haloes  in   various
cosmologies,  including power  law   and    open models.   For     the
$\Lambda$CDM model we discuss,  their favorite parameters are $f=0.01$
and $k  = 3.4\times10^3$, and these  provide a reasonable reproduction
of the median $\cvir (M)$ relationship at $a=1$ in our simulations.

Although useful in its ability to provide  the correct $z=0$ relation,
this   model fails to reproduce  the  observed  $z$  evolution of halo
concentrations (\S~\ref{sec:redshift}).   If  the  value  of  $f$   is
adjusted to  the  unphysically small value  of $f=10^{-15}$  then  the
redshift behavior is becomes similar to that reported in
\S~\ref{sec:redshift}.  However, this value of $f$ pushes the relevant
region of  the  power spectrum  to  extremely small masses, where  the
effective slope  is extremely   flat, $\alpha  \lsim  0.05$,  so the
implied $\cvir(M)$ dependence becomes shallower than what is observed.
In  \S~\ref{sec:toy}, we present a  revised toy model which, using the
same number of free parameters, reproduces  the full observed behavior
of $\cvir(M,z)$.

\section{The halo finder}

\def\pmb#1{\setbox0=\hbox{#1}%
 \kern-.025em\copy0\kern-\wd0
 \kern.05em\copy0\kern-\wd0
 \kern-.025em\raise.0433em\box0}

\def\vv{\pmb{$v$}}
\def\vq{\pmb{$q$}}
\def\vx{\pmb{$x$}}
\def\vr{\pmb{$r$}}

Most commonly used halo finders, which  work either by the location of
overdensities in a spatial window  of fixed shape (usually  spherical)
or by friends-of-friends algorithms, do  not account for haloes within
haloes.   Since  our projects   specifically  address the question  of
substructure, we   have  been obliged   to devise   a halo finder  and
classification algorithm suited for this purpose.

If one were  only interested in  distinct,  virialized objects, haloes
would be  easily identified --- there is  little confusion as to where
one halo ends and another  begins  because the  physical extent of  an
object is determined by the virial overdensity criterion.  However, we
are  interested in objects  within the virial  radius of large haloes,
therefore certain  ambiguities  arise.   How close  must   two density
maxima be in order  for them to  represent a single object?   How does
one differentiate substructure from a collision in progress?  How does
one assign mass to haloes and subhaloes appropriately?

We have found a  solution to these  problems by assigning to each halo
two length scales --- an inner radius, $\Rs$,  and and an outer radius
$\Rvir$.  We do so by modelling the density profile of each halo using
Eq.  \ref{eqt:nfw}.  The  virial radius  $\rvir$  determines each halo
mass and radial  extent,    and  $\rs$ determines when    two  density
maxima/haloes  should be combined into  one. The details are described
below.

Because the modelling process  requires  fitting a density profile  to
each halo, we  attempt  to find  only haloes with  more  than $\Npmin$
particles within their  virial radii.   If  $\mp$ is the mass  of each
particle,  this means the minimal  virial mass of haloes identified is
$\Mmin = \Npmin \times \mp$.   Equivalently, using Eq. \ref{eqt:mvir},
we have  a  minimum virial radius  $\rvir^{\rm   min}$.  The value  of
$\Npmin$ is  the first free    parameter of this  algorithm.  We   use
$\Npmin = 50$.

Our density maxima finding routine is based on the BDM (\citeNP{kh})
algorithm.  We  outline our   procedure  below, including  our precise
methodology which is generalizable for  any simulation parameters  and
our detailed procedure for defining and classifying haloes.

\begin{enumerate}

\item 
\label{item:one}
\ni
We  construct density field  values  by a Cloud-in-Cell (CIC)  process
\cite{hockney:81} on the  largest grid of  the  simulation $\Delta L$,
and rank the particles according to their  local density as determined
on this grid.

We then   search  for the possible  halo  centres,  using two sets  of
smoothing spheres;  one,  with a small radius,  $\rspo$,   in order to
locate  of tight, small clumps;  and the other,  with a larger radius,
$\rspt$, in order to locate the  centres of haloes with diffuse cores.
The smaller radius is $\rspo = \alpha \, \fres$,  where $\fres$ is the
highest force   resolution in the  simulation and  $\alpha$ is  a free
parameter of  order  unity.  We use $\alpha   = 2.$ The second  set of
spheres have $\rspt = \Rvirmin$.

For each set  of spheres, we  take from  the ranked list  the particle
with the highest local density and place  a sphere about its location.
A  second sphere is  placed  about the next  particle  in the list not
contained in the first sphere.  The  process is continued until all of
particles are  contained within at  least one sphere.  Because  we are
only interested in centres of haloes  more massive than $\Mvirmin$, we
discard each sphere with fewer  than a set number  of particles.   The
minimum number  of particles required for  a kept sphere is determined
separately for each radius.

For  the  $\rspo$  spheres, we  use   the following  conservative halo
density profile:
\begin{eqnarray} 
  \rho(r) = \left\{\begin{array}{ll} 
                C/r^{2.5}_{\rm sp1} &\mbox{$r<r_{\rm sp1}$} \\ 
                C/r^{2.5}  &\mbox{$r>r_{\rm sp1}$},
                \end{array} 
        \right.
\end{eqnarray}
(where  $C$  is  determined my  fixing   the minimum halo mass   to be
$\Mvirmin$), in order to estimate the minimum core number of particles
within $\rspo$:
\begin{eqnarray}
 \Nspo = \frac{\Npmin }
       {1  + 6[(\Rvirmin/\rsp)^{1/2} - 1]}.
\end{eqnarray}
For the $z=0$  timestep of the  $60 h^{-1}$Mpc  simulation we analyze,
$\Nspo = 3.9 \rightarrow 3$.  Spheres  of size $\rspo$ with fewer than
$\Nspo$ particles are  discarded.     Similarly, all of  the   $\rspt$
spheres  containing  fewer   than   $\Nspt  =  \Npmin$  particles  are
discarded.

The final list of candidate   halo centres is  made up  of all of  the
(small)  $\rspo$ spheres, together  with  each of the $\rspt$  spheres
that \textit{does not} contain an $\rspo$ sphere.

\item  For each sphere of radius $r_{sp} = \rspo$ or $\rspt$, whichever
applies, we use the  particle distribution to find  the centre of mass
and iterate  until convergence.    We  repeat the procedure   using  a
smaller  radius,  $r=r_i$,  where $r_i=r_{sp}/2^{i   \over   2}$.   We
continue this method until $r_i = r_L$, where  $r_L$ is defined by the
criterion $r_{L} > f_{\rm res} > r_{L+1}$, or until reduction leads to
an empty sphere.

\item We unify the spheres whose centres are
within $r_{L}$ of each other.   The unification is performed by making
a  density weighted   guess for a  common centre   of  mass, and  then
iterating to find a centre of mass for  the unified object by counting
particles.  The size of sphere used to determine the centre of mass is
the smallest radius that will allow the new sphere to entirely contain
both candidate halo spheres.

\item For each candidate halo centre we step out in radial shells of 
$1   \hkpc$, counting enclosed particles, in   order to find the outer
radius of the halo: $R_h \,=$  min($\rvir,\,R_t$).  The radius $\rvir$
is the virial radius, and $R_t$ is a ``truncation'' radius, defined as
the  radius  $(< \rvir)$  in which a  rise in   (spherical) density is
detected  ($d   \log\rho/d \log  r  >  0$).  This    is our method for
estimating when a different halo  starts to  overlap with the  current
halo and is important for haloes  in crowded regions.  We estimate the
significance of a measured upturn  using the Poisson noise  associated
with  the number of particles in  the radial bins considered.  Only if
the signal to noise of the upturn  is larger than $\sigma_{R_t}$ do we
define  a truncation  radius.  The  value of  $\sigma_{R_t}$ is a free
parameter.  We  use   $\sigma_{R_t} =  1.5$.~\footnote{The choice  was
motivated by several tests using mock catalogues of haloes in clusters
designed to determine how varying  $\sigma_{R_t}$ affects our  ability
to fit  the density profiles of  subhaloes.  Although our results were
not  strongly dependent  on this choice,  we did  obtain the best fits
using $\sigma_{R_t} = 1.5$.}

\item Among the halo candidates for which we have found an $\rvir$,
we discard those with $N_{\rm vir}<\Npmin$, where $N_{\rm vir}$ is the
number of particles  within $\rvir$.   Among  the halo candidates  for
which we have   found a rise   in spherical density, we discard  those
which contain less  than $N_{\rm \Rt}^{min}$  particles, where $N_{\rm
\Rt}^{min} = \Npmin$ if $\Rt > \Rvirmin$, otherwise
\begin{eqnarray}
	 N_{\rm \Rt}^{min} = 
         N_{\rm p}^{min}{\left(\frac{\Rt}{\Rvirmin}\right)}.
\end{eqnarray}
The above constraint follows from an extrapolation of the minimum mass
halo using an isothermal profile $\rho(r) \propto 1/r^2$.

\item We model the density profile of each halo using the NFW form
(Eq~\ref{eqt:nfw}) and  determine   the best fit  $\rs$  and  $\rho_s$
values,  which determine $\rvir$ and   $\Mvir$.  The fitting procedure
uses logarithmically spaced radial  bins from max($2f_{\rm res},  0.02
\times {\rm min}(\rvir,\Rt)$) out to $R_h$.   If any bins are empty we
decrease the  number of bins  by one until  all bins are full.  If the
number of  bins  is reduced  below  3 we discard the  halo  as a local
perturbation.

The  fit takes into account  the Poisson error  in each bin due to the
finite number of particles, and we obtain errors on the fit parameters
($\sigma_{\rs}$ and $\sigma_{\rho_s}$)  using the covariance matrix in
the fit routine.   The errors on the fit  parameters can be translated
easily  into errors for   $\rvir$ ($\sigma_{\rvir}$) and the estimated
NFW mass of each halo, $\Mvir$ ($\sigma_M$).

\item We unify haloes which overlap in $R_{\rm s}$.
Our  criterion is met if  two (or more)  halo centres have $R_{\rm s}$
radii  which overlap  with  each other while   at the same time having
velocities which allow them to be bound to the common system.  If such
a case  occurs, then along with the  individual halo NFW  fits, we fit
another  NFW  profile about  the common   centre  of mass  of  the two
combined haloes  and  decide  whether the candidate-united-haloes  are
bound/unbound to  the common NFW  fit using the radial escape velocity
determined using the common  NFW profile (see  below).  If both haloes
are bound we combine the two haloes into one, and  keep the common fit
for the characteristic parameters.   If at least one  is not bound, we
do not combine the haloes.

An exception  to this unifying criterion  occurs if  the fit errors on
$R_{\rm s}$ are large ($\sigma_{R_{\rm s}}/R_{\rm s} > 1$), we replace
$R_{\rm s} \longrightarrow \rm{min}(R_{\rm  s}, R_t)$.  In addition if
the $R_{\rm s}$ of a  halo obeys $R_{\rm s} >  \Rvirmin$ then we relax
the strict combining of overlapping haloes.  This case, which we refer
to as the ``cD'' halo case, is discussed below (see
\ref{item:cd_case}.)

\item For each halo, we remove all unbound particles 
before we  obtain the final fits.  We  loop over  all particles within
the halo and declare a particle at a distance $r$ from the centre of a
halo to  be unbound if   its velocity relative to   the centre of mass
velocity  of  the halo    obeys  $ v   > \sqrt{2\left|\Phi_{\sss   \rm
NFW}(r)\right|  \,},$   where the  radial  potential   for NFW density
profile is   given by\footnote{Note that this  potential  is {\it not}
necessarily  the   physical   gravitational potential  at    the  halo
location. For a subhalo, for example, the host background potential is
{\it not} included. }
\begin{eqnarray}
   \Phi_{\sss \rm NFW}(r) = 
  -4\pi G \rho_{\rm s} {R_{\rm s}}^2\left[ \frac{\log(1 + x)}{x}\right].
\end{eqnarray}
 
After removal, we construct a new density profile and find new NFW fit
parameters.  The procedure is   repeated until  the number  of unbound
particles becomes    $<  1  \%$   of the   bound  particles or   until
$\Mvir^{\sss \rm NFW} < \Mmin$, in which case the halo is discarded.

An exception to  this removal scheme  occurs if two haloes  lie within
the virial radius of  each other and the  ratio of their masses are at
least $0.75$.  We define haloes in this situation  to be a ``partner''
pair.  For each halo in this situation, we take not only its potential
into account, but also that of its partner.

\item 
\label{item:cd_case}
\ni An interesting case of subhalo structure, 
which would otherwise be excluded from our finding algorithms, is that
of one or more density peaks close to the centre of a  large halo.  We
shall refer to these   inner density peaks   as cD cases.  If  a halo,
after its unbound particles   have been removed, obeys the   following
criteria, it  is a candidate for containing  cD haloes: a) the NFW fit
has a standard GoF $<$  0.1,  b) the halo is  a  host of at least  one
subhalo, and c) the halo is ``large'', with $\rs > \Rvirmin$.

We identify the potential centres of cD  haloes by searching the bound
particle distribution within  $\rs$ of each  candidate cD halo using a
CIC process on a fine grid ($\rspo$ = $\alpha \fres$).  We discard all
candidate density  peaks    with   local densities less     than   the
extrapolated minimum density (above the background density) within the
core region of our smallest halo (see item~\ref{item:one}.)

For each density  maxima located farther than  $\rspo$ from the centre
of the candidate host halo,  we find $\Rt$ and  fit a NFW profile with
iterative  unbound particle  removal.  These are   our cD haloes.   cD
haloes are  discarded if their extrapolated virial  mass is lower than
our minimal mass halo.

\end{enumerate}

Because we have a strict mass  limit $\Mvirmin$ = $\Npmin \times m_p$,
we expect our  halo  finder  to be   somewhat incomplete  just   above
$\Mvirmin$.  We   have  also checked our   completeness  in  two ways.
First, we used a separate BDM halo finder that does not attempt to fit
profiles and  does not   demand  the unification   of haloes within  a
specified radius.  It does,  however,   unify haloes that have   equal
velocities within $15\%$ as long as they have centres within $\sim 150
\hkpc$    (see~\citeNP{kh}).   This  procedure    allows  a   complete
identification of  DM haloes down  to much lower particle numbers than
our own.  In order to check our results we have  assigned to each halo
in the catalog produced by the other finder  a typical $\rs$ given its
mass, and checked the  returned halo list  for consistency against our
catalogue from  the same simulation box.   For $N_{\rm particles} \sim
150$, we estimate $\sim 80\%$ completeness  and for $N_{\rm particles}
\sim 500$  we obtain $\gsim 95\%$ completeness.   A second, and almost
identical completeness  determination    is  obtained   by   carefully
analyzing the roll-over in our observed mass function
\cite{sigad:99}.
We attribute our incompleteness    for  small masses to    our fitting
procedure, and errors associated with this process.

\section{REDUCING TF SCATTER}

As discussed in \S 6, we find  for distinct haloes a $1\sigma$ scatter
at fixed  $\Mvir$  of  $\Delta  \Vmax /\Vmax   \simeq 0.12$, which  is
between 1.3  and 4   times the  range  of the  reported intrinsic   TF
scatter.  If these $\Lambda$CDM haloes are to host galaxies like those
observed, this excessive scatter must be  reduced.  The translation of
the halo virial mass into a disc  luminosity, and of the original halo
$\Vmax$  into a final  observed disc velocity, should somehow decrease
the scatter.  Following is a qualitative analysis of how this can come
about  in a natural  way. The idea  is that for a  fixed halo mass and
spin,  a higher $\vmax$ should   induce further  gas contraction  into
smaller radii, and therefore  higher gas density, star-formation  rate
and luminosity.   This  can  be shown   in a  little more   detail, as
follows.
 
The size of  the exponential disc, $\rd$, that  forms by a dissipative
contraction  of gas inside  a given dark-matter  halo can be estimated
under the adiabatic baryonic-infall  approximation.  We showed in \S 2
(see Equation~\ref{eqt:rd}) that   $\rd$ is a  decreasing  function of
$\cvir$  for haloes of a fixed  virial mass and  spin, as  long as the
disc mass is a constant fraction of $\Mvir$.  For a  typical case of a
$\vvir  =  200$   km/s halo,   we  demonstrated  that,   in the  range
encompassing 68\% of $\cvir$ for such haloes ($\cvir$:$9.2 \rightarrow
21.3$),  the  corresponding   spread  in   disc  sizes is   $R_{d}:4.0
\rightarrow  2.6\hkpc$.   Across  this range,   an effective power-law
approximation would therefore be  $\rd \propto \cvir^{-0.5}$.   If (a)
the gas density  in the disc scales  like $\rho \propto \rd^{-2}$, (b)
the star formation rate obeys a typical Schmidt law, $\dot\rho \propto
\rho^{1.5}$,  and (c) the  luminosity scales like  $L \propto \dot\rho
\rd^2$, then the luminosity  at a given mass  depends on $\cvir$ as $L
\propto   \rd^{-1} \propto \cvir^{0.5}$.  Since  $\vmax$  (for a fixed
$\vvir$) is also a  monotonic function of $\cvir$ (Eq.~\ref{eqt:vmax};
the effective power-law approximation across the  68\% range is $\vmax
\propto  \cvir^{0.27}$),   we  have  obtained  a  positive correlation
between $L/\mvir$ and the deviation of $\vmax$ from the median $\vmax$
at a given $\mvir$.  Ignoring, for the  moment, any difference between
the $\vmax$ of the  original halo and that of  the disc, the  obtained
correlation  would correspond to a reduced  scatter in $L$ by a factor
larger than 2, as required.  The effect of the  spread in spins on the
TF  scatter  is expected to be   reduced for similar  reasons, namely,
because the luminosity and the  maximum velocity are both expected  to
correlate with the spin in the same sense.

More detailed modelling, which takes into  account how $\vmax$ changes
as the baryons  fall  in, will be   needed to test  this hypothesis in
detail.

\bibliographystyle{mnras}
\bibliography{mnrasmnemonic,paper1}
                                                
\end{document}